\documentclass[prb,english,notitlepage,twocolumn,superscriptaddress]{revtex4-1}
\usepackage{amssymb}
\usepackage{hyperref}
\usepackage{graphicx}
\usepackage{amsmath}
\usepackage{color}
\usepackage{float}
\usepackage{subfigure}
\usepackage{placeins}
\usepackage{bm}

\makeatletter
\renewcommand*\env@matrix[1][\arraystretch]{%
  \edef\arraystretch{#1}%
  \hskip -\arraycolsep
  \let\@ifnextchar\new@ifnextchar
  \array{*\c@MaxMatrixCols c}}
\makeatother

\newcommand{\COMMENTED}[1]{}

\begin{document}

\title{Weyl Nodal-Ring Semimetallic Behavior and Topological Superconductivity in Crystalline Forms of Su-Schrieffer-Heeger Chains}
\author{Peter Rosenberg}
\affiliation{Département de Physique \& Institut Quantique, Université de Sherbrooke, Québec, Canada J1K 2R1}
\affiliation{National High Magnetic Field Laboratory and Department of Physics, Florida State University, Tallahassee, Florida 32306, USA}
\author{Efstratios Manousakis}
\affiliation{National High Magnetic Field Laboratory and Department of Physics, Florida State University, Tallahassee, Florida 32306, USA}
\affiliation{Department of Physics, National and Kapodistrian University of Athens, Panepistimioupolis, Zografos, 157 84 Athens, Greece}
\begin{abstract}
We consider a three-dimensional model of coupled Su-Schrieffer-Heeger (SSH) chains. The analytically soluble model discussed here reliably reproduces
the features of the band structure of crystalline polyacetylene as obtained from density-functional theory.
We show that when a certain inter-chain coupling is sufficiently increased, the system develops a ring of Weyl nodes.
We argue that such an increase could be achieved experimentally by intercalation or extreme pressure.
With the addition of a simple intra-orbital pairing term we find that the system supports an exotic
superconducting state with drumhead surface states and
annular Majorana states localized on the surface.
In addition to suggesting a novel real material realization of a nodal ring semimetal and possibly topological superconductivity, our results provide a new perspective on the SSH model,
demonstrating that a simple extension of this broadly-impacting model  can once again provide fundamental insights on the topological behavior of
condensed matter systems.
\end{abstract}

\maketitle

\section{Introduction}

Since its introduction over four decades ago, the Su-Schrieffer-Heeger (SSH) model \cite{SSH1979} has served as a beautiful and rather simply understood
example of emergent quasiparticles, qualitatively different from those in the non-interacting system. 
The model was inspired by quasi-one-dimensional materials like polyacetylene, and despite its apparent simplicity, has proven to be a tremendously rich description 
that captures a variety of fascinating phenomena, including solitons, topological transitions, edge states, and charge fractionalization \cite{SSH_RMP,Rice1982}.
The SSH model also provided one of the earliest examples of a non-trivial one-dimensional Berry phase, known as the Zak phase \cite{Zak1989}, in a condensed matter system.

These early notions of topology, first explored within the context of the SSH model and in the quantum Hall effect, have become a central focus of modern condensed matter physics. 
This renewed interest in the role of topology in condensed matter systems was primarily motivated by the discovery of topological insulators \cite{Hasan2010}. More recently a new class of topological 
materials has been discovered, the Dirac and Weyl semimetals \cite{Armitage2018,Liu2014,Lu2015,Lv2015,Xu2015,Huang2016,Jiang2017}. These materials are defined by the presence of a set of topologically protected band touching points, which lead to unique transport properties. The set of proposed topological semimetals has since been expanded to include systems with lines or rings of degeneracies 
in their band structures, which have been termed nodal line or nodal ring semimetals \cite{burkov_topological_2011,kim_dirac_2015,chen_nanostructured_2015,fang_topological_2016}. Several examples of 
these systems have already been proposed or reported in real materials \cite{yu_topological_2015,bian_drumhead_2016,Bian2016}.
These novel materials hold the promise of many impactful applications, making a reliable description of their exotic behaviors a priority of condensed matter physics.

In this work we provide a new perspective on the well-studied SSH model, illustrating that this relatively simple description, which yielded some of the earliest insights into the topology of condensed matter systems, 
can be adapted straightforwardly to describe the physics of topological nodal semimetals. We present a three dimensional analytically soluble model, inspired by the structure of crystalline polyacetylene, consisting of coupled SSH chains.
We show that with an appropriate choice of the hopping parameters it
reliably reproduces the features of the band structure of crystalline polyacetylene, as obtained from density-functional theory.
In addition, we demonstrate that the band structure of this model, when
a diagonal hopping parameter is increased,
contains a Weyl nodal ring. It is conceivable that this parameter regime
could be experimentally accessed by application of extreme pressure
on crystalline polyacetylene. We also find that, with the addition of simple
pairing interactions, the model supports a topological superconducting phase characterized by the presence of
surface Majorana fermions. While our model could be realized experimentally in the context of conjugated polymers, it is just as relevant to the case of cold atoms, where there has been remarkable progress in the realization of topological lattice models \cite{Atala2013,Cooper2019}, including Weyl semimetals \cite{wang_realization_2021} and nodal ring semimetals \cite{song_observation_2019}.     

The observation of superconductivity in topological systems has brought considerable attention
to the problem of the interplay of strong correlations and pairing with topology, including in the context of Dirac and
Weyl semimetals \cite{Balents_2012,Cho2012,Burkov_2015,Haldane_2018,kobayashi_topological_2015,Alidoust2017,2DWeyl_AFQMC}, as well as nodal line semimetals \cite{nandkishore_weyl_2016,wang_topological_2017,fu_transport_2020}, which may support exotic superconducting phases. 
Many models are known to support topological superconductivity, typically characterized by the presence of Majorana surface or edge states. 
Most theoretical descriptions of these exotic superconducting states assume the presence of $p$-wave or higher angular momentum pairing terms.
However, it was recently noted that topological superconducting states can emerge from conventional s-wave pairing interactions in Weyl systems. 
In these systems each Weyl node is split by the interaction into a pair of Bogoliubov-Weyl  \cite{Balents_2012,Burkov_2015,faraei_induced_2019}
nodes and Majorana states appear at the boundaries. This behavior has also been recently explored in nodal ring semimetals \cite{fu_transport_2020} and does
 indeed emerge in our model with the addition of an inter-orbital pairing term.

\section{Model}

\label{sec:latt_ham}
We begin with a simplified TB model of the $P2_1{/}a$ structure of crystalline \textit{trans}-polyacetylene, as depicted in Fig.~2 of Ref.~\cite{Vogl1990}.  
Our model keeps all the relevant elements and symmetry aspects of the real structure, but is simple enough to permit an exact solution. This compromise is 
made in order to project out and provide a clear understanding of the new physics of this system. Despite its apparent simplicity, the model shows
remarkable agreement with the DFT calculation.

\begin{figure}
    \begin{center}
         \includegraphics[width=0.85\columnwidth]{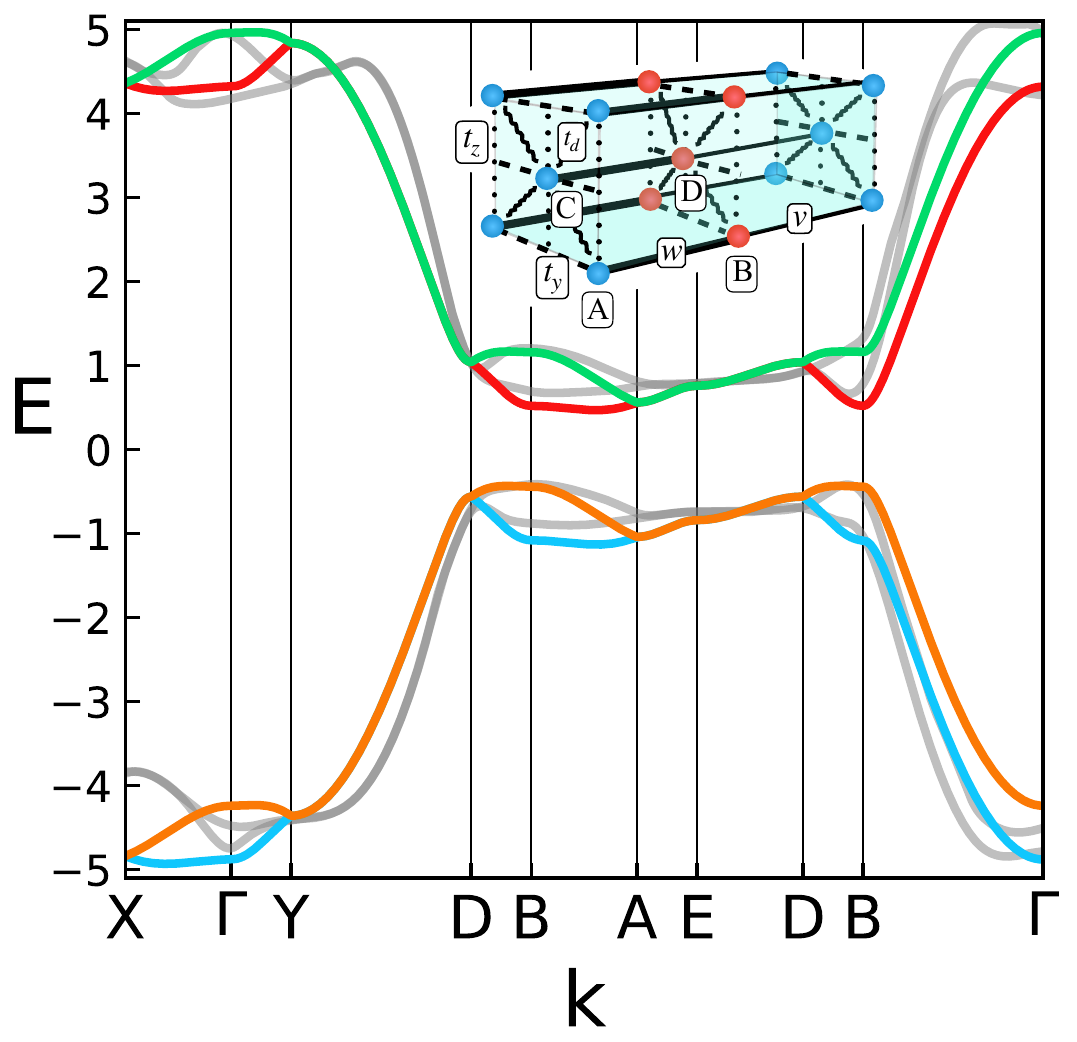}
    \end{center}
    \caption{           \label{fig1}
Tight-binding versus DFT band structure. We show the band structure obtained from our lattice model, compared to
density functional theory results using the experimentally measured structural parameters \cite{Vogl1990}. We include
an overall shift to the DFT results to restore particle-hole symmetry about $E=0$. The inset shows the unit cell with relevant
hopping elements.}
\end{figure}

In the inset of Fig.~\ref{fig1} we present the unit cell and hopping elements for our model, which provides a qualitative description of the important features of
the band structure of the $P2_1{/}a$  structure of crystalline \textit{trans}-polyacetylene.
The dimerized SSH chains are along the $x$-axis. The red and blue spheres represent carbon atoms whose color corresponds to their relative position along the dimerized chain (A or B sublattice). 
The dimerization leads to two different hopping matrix elements, denoted here as $w$ and $v$,
along the chain. These SSH chains run parallel to each other, forming the unit cell depicted in the inset of Fig.~\ref{fig1}. Though depicted here as collinear, in polyacetylene the carbon atoms do not lie along a straight line,
they form bonds of $120^{\circ}$ to facilitate the $sp^2$ hybridization of the 2$s$ and two of the three 2$p$ orbitals that form the bonds of the carbon atom
with its nearest neighboring carbon atom and a hydrogen atom (neglected in the
structure depicted in Fig.~\ref{fig1}).
The hopping of the third $2p$ (call it $2p_z$) electron
of the carbon atom forms the bands near the Fermi level. 
While the physical presence of the hydrogen atoms along each chain and their relative orientation is important in determining the effective
interchain hopping matrix elements, these atoms are
also abstracted out of our drawing for simplicity.

In addition to the hoppings $w$ and $v$, we consider a carbon to carbon interchain hopping matrix element
along the $y$ ($t_y$) and the $z$ ($t_z$) directions. We also include a hopping matrix element along the diagonal $(0,1,1)$ direction $(t_d)$.
The most general version of the model includes an additional hopping along the $(1,1,1)$ direction $(t^\prime_d)$ (depicted in Fig.~\ref{fig2}(a)),
but we first consider the limit $t^\prime_d \rightarrow 0$, corresponding to the unit cell shown in Fig.~\ref{fig1}.
The unit cell is doubled along the $x$ direction because of the dimerization along that direction; however, the interchain hopping matrix elements are 
independent of whether the hopping is between carbon atoms on the A sublattice or the B sublattice.

The TB Hamiltonian of the structure is given in $k$-space as $\hat H_0 = \sum_\mathbf{k} c^\dagger_{\mathbf{k}} \mathcal{H}_0(\mathbf{k}) c_{\mathbf{k}}$, with,
\begin{eqnarray}
  \mathcal{H}_0(\mathbf{k}) =  \left ( \begin{array}{cccc} 
e({\bf  k})  & S & V  & V^\prime \\
S^* & e({\bf  k}) & V^\prime & V \\
V & V^\prime & e({\bf  k}) & S \\
  V^\prime & V & S^* & e({\bf  k})\end{array} \right ),
  \label{eq:H0_nrs}
\end{eqnarray}
where $c^\dagger_\mathbf{k} =\begin{pmatrix} c^{\dagger(\textmd{A})}_\mathbf{k} & c^{\dagger(\textmd{B})}_\mathbf{k} & c^{\dagger(\textmd{C})}_\mathbf{k} & c^{\dagger(\textmd{D})}_\mathbf{k} \end{pmatrix}$ is the set of creation
operators for electrons in the $2p_z$ state on the A, B, C and D atoms (see Fig.~\ref{fig1} for labels), and
\begin{eqnarray}
  e({\bf k}) &=& -2 t_y \cos(k_y a_y) - 2 t_z \cos(k_z a_z),\\
  S &=& X + i Y, \\
  X&=&(v+w) \cos(k_x a_x/2), \\
  Y &=& (v-w) \sin(k_xa_x/2), \\
  V &=& -4 t_d \cos(k_ya_y/2)\cos(k_za_z/2), \\
  V^\prime &=& -8 t^\prime_d \cos(k_xa_x/2)\cos(k_ya_y/2)\cos(k_za_z/2).
\end{eqnarray}
All hopping parameters are in units of eV.

We first consider the case of $t^\prime_d=0$.
Our model can be diagonalized exactly and has the following set of bands:
\begin{align}
E^\pm_1(\mathbf{k}) & = -2t_y\cos(k_ya_y)-2t_z\cos(k_za_z) \notag \\ 
					&\pm \sqrt{X^2+Y^2+V^2+2 \vert V \vert\sqrt{X^2+Y^2}} \notag \\
E^\pm_2(\mathbf{k}) & = -2t_y\cos(k_ya_y)-2t_z\cos(k_za_z) \notag \\
					&\pm \sqrt{X^2+Y^2+V^2-2 \vert V \vert\sqrt{X^2+Y^2}}.
					\label{eq:bands}
\end{align}
In Fig.~\ref{fig1} we compare the bands from our TB model to the band structure obtained via DFT using the 
experimentally determined structural parameters \cite{Vogl1990}. We observe that the features of the DFT band structure 
(gray curves in the figure) are well reproduced.

\section{Nodal ring semimetal and drumhead states}

\begin{figure}
    \begin{center}
           \includegraphics[width=\columnwidth]{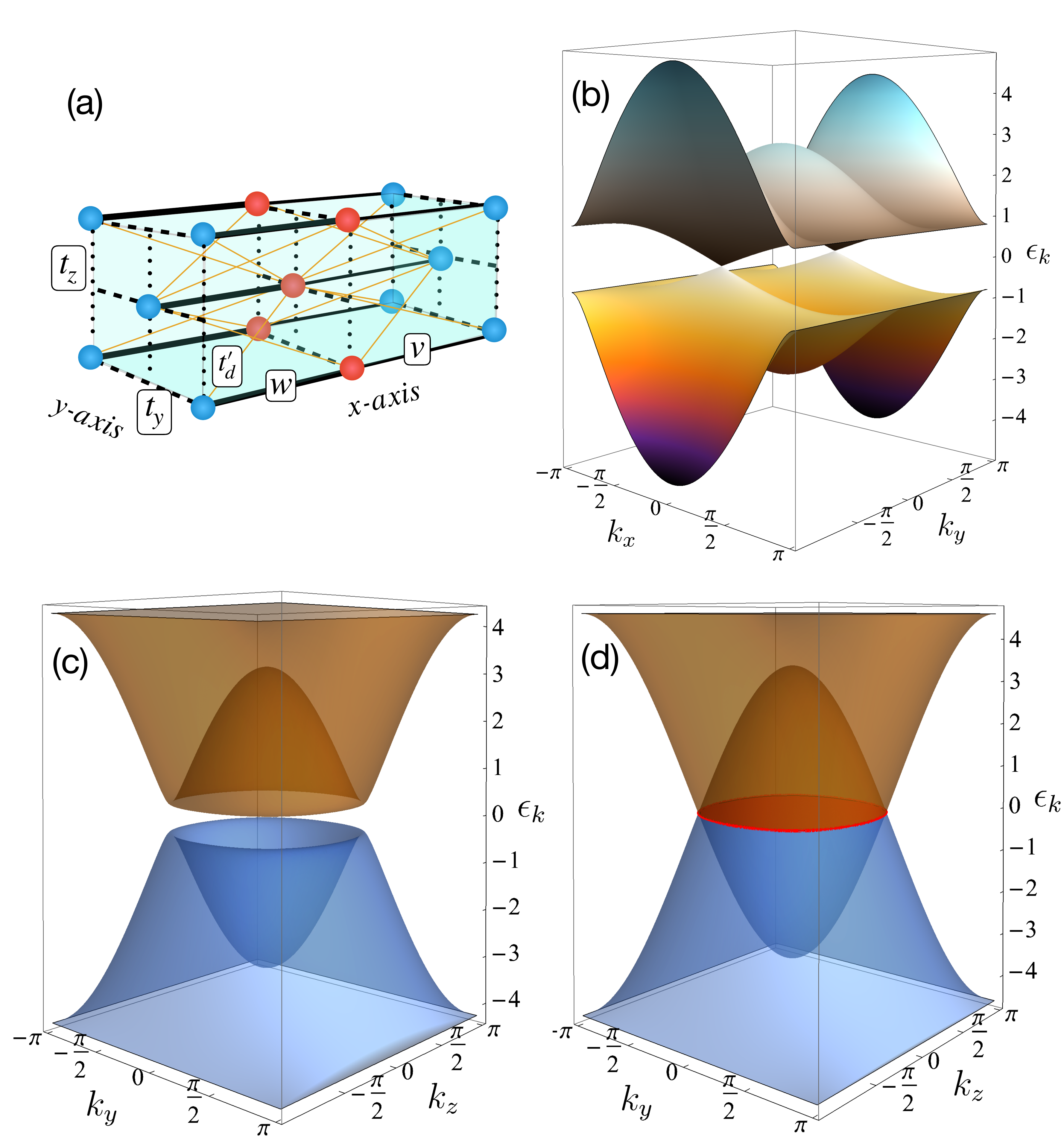}
    \end{center}
    \caption{           \label{fig2}
(a) Unit cell of modified lattice TB model. (b) Nodal points in $k_x$-$k_y$ plane at $k_z=\pi/4$.
    (c),(d) Band structure in the $k_y$-$k_z$ plane at different values of $k_x$. (c) $k_x=\pi/4$. (d) $k_x=0$. 
    The nodal line at $k_x=0$, which bounds the drumhead surface
    state (light red), is highlighted in red. The model parameters are $v=2.7$, $w=1.9$, $t_d^\prime=1.0$, $t_d=t_y=t_z=0$,
    and we take $a_x=a_y=a_z=1$.}
\end{figure}

\begin{figure}
    \begin{center}
      \includegraphics[width=\columnwidth]{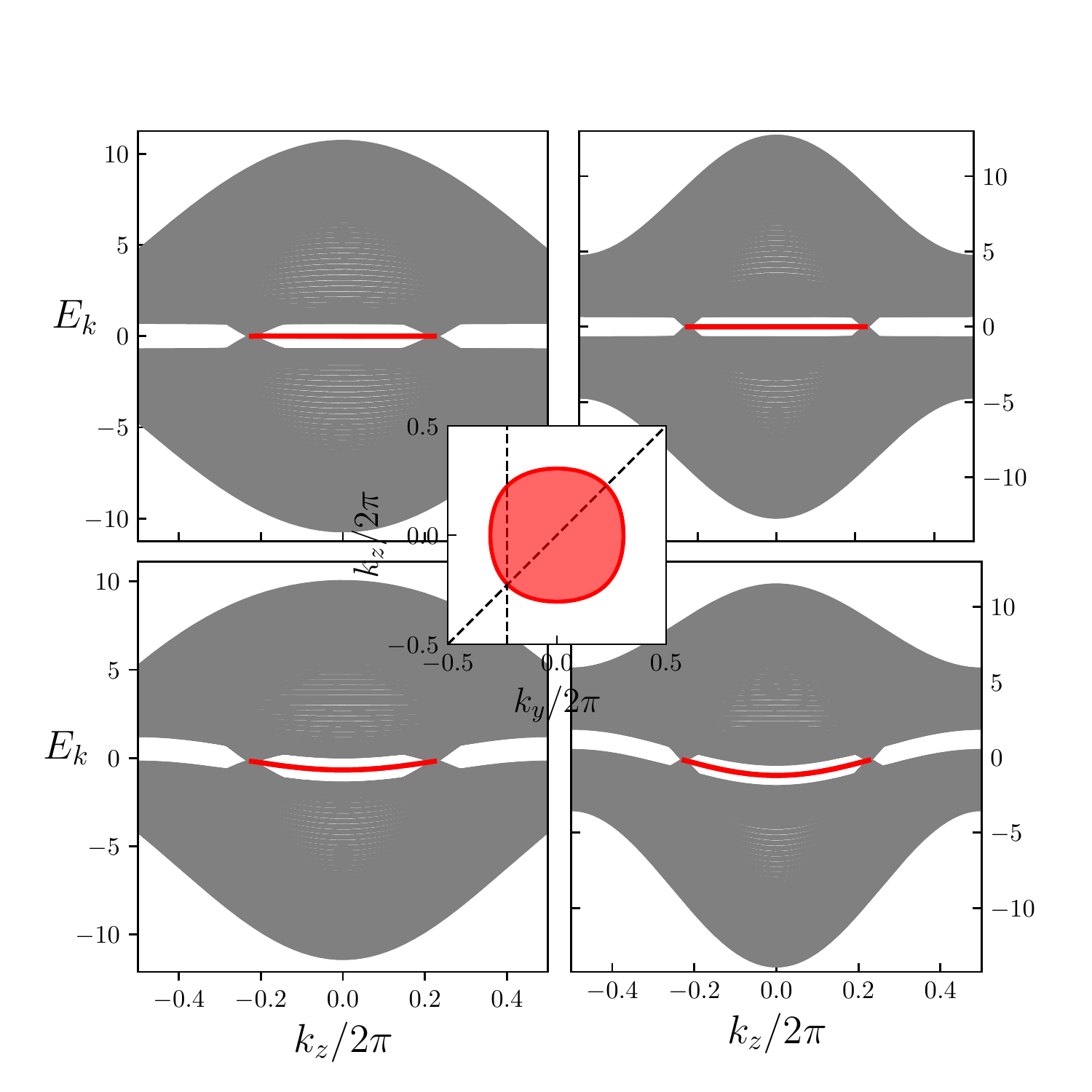}
    \end{center}
    \caption{Surface states versus $k_z$ for a slab open along $x$.
      The surface state is highlighted in red.
      The top two panels show the edge spectrum with $t_y = t_z = 0$, and the bottom panels show the surface spectrum for $t_y = t_z = 0.3$
      (The remaining parameters are the same as in Fig.~\ref{fig2}).
The left column is along the cut $k_y/2\pi = -0.228$, and the right column is along $k_y = k_z$. (Inset) Nodal ring
in the $k_y$-$k_z$ plane and drumhead state in the shaded red region.
The dashed lines correspond to the cuts used for the two columns. 
 \label{fig3}}
\end{figure}

We now consider a related, analytically soluble model, by taking the limit of $t_d \rightarrow 0$ while $t^\prime_d\neq0$, and illustrate 
that this simple modification of the previous model can realize a 3D Weyl nodal ring semimetal. We note that for the structure depicted in 
Fig.~\ref{fig2} (i.e., $P2_1{/}a$) this requires the second closest neighbor hopping (diagonal, blue to red) to be dominant over the closest neighbor hopping 
(diagonal, blue to blue), which may be difficult to achieve experimentally.
However, the Hamiltonian corresponding to this case has an identical form to the Hamiltonian for the $P2_1{/}n$ structure
(see Appendix \ref{sec:AppendixA}), with the second closest neighbor hopping, $t^\prime_d$, becoming a closest neighbor hopping.
This Hamiltonian, with larger interchain-hopping matrix elements, may be realized by the application of extreme pressure, which should make the
system more three dimensional, as opposed to the currently experimentally accessible structure that has a
quasi-1D nature. The application of high pressure  
may stabilize one or the other structure, both of which support nodal ring states. 
Additionally, while we focus on the limit of $t_d\rightarrow0$ for demonstration,
because it permits an analytic solution, we observe that the nodal ring state exists across a broad parameter regime, including for the
case of both finite $t_d$ and $t^\prime_d$, and is also robust with respect to the magnitudes of $t_y$ and $t_z$ (see Fig.~\ref{fig3}). 
 
 The eigenvalues in this case ($t_d=0$, $t^\prime_d\neq 0$)  are the following four bands:
\begin{eqnarray}
  E^{+}_{\pm}({\bf k}) &=& e({\bf k}) \pm \lambda_+,\\
  E^{-}_{\pm}({\bf k}) &=& e({\bf k}) \pm \lambda_-,\\
  \lambda_{\pm} &=& \sqrt{(|X| \pm |V^\prime|)^2 + Y^2}.
\end{eqnarray}
A ring of Weyl nodes is formed by the $E^{-}_{\pm}$ bands at the set of momenta satisfying the equations,
$Y=0, \hskip 0.2 in |X| = |V^\prime|$.
The solutions to these equations are,
\begin{eqnarray}
  k_x=0, \hskip 0.1 in \left\vert \cos\left({{k_za_z}\over 2}\right)\right\vert
  \left|\cos\left({{k_ya_y} \over 2}\right)\right| = {{v+w} \over {8t^\prime_d}},
  \label{nodal_points}
  \end{eqnarray}
which specify a nodal ring, provided $r\equiv (v+w)/(8t^\prime_d) \le 1$.
We show an example of the nodal ring in Appendix \ref{sec:AppendixB}.

In Fig.~\ref{fig2}(c)-(d) we illustrate the emergence of the nodal ring in the bulk band
structure as the value of $k_x$ approaches zero.
For clarity we show only $E^{-}_{\pm}$, and note that the $E^{+}_{\pm}$ bands
do not participate in the formation of the nodal ring.
When we fix $k_z$ (or $k_y$) and plot the band-structure 
in the $k_x$-$k_y$ (or $k_x$-$k_z$) plane we observe a pair of Weyl nodes (see Fig.~\ref{fig2}(b)), the full set of which, obtained by varying $k_y$ and $k_z$, forms the nodal line.

When the system is finite in the $x$ direction, a doubly degenerate drumhead 
surface state bounded by the nodal ring forms \cite{chan_ca_2016}. 
This surface state is depicted in red in Fig.~\ref{fig2}(d).
The surface spectrum is plotted versus $k_z$ in Fig.~\ref{fig3}
for two different choices of the hopping parameter $t_y$ and
$t_z$ and for two values of $k_y$. The drumhead surface state is again highlighted in red. Notice that when $t_y$ and $t_z$ are non-zero, the drumhead state
becomes dispersive.

\section{Berry Phase}
The eigenstates of the $\lambda_-({\bf k})$ eigenvalue, which correspond to
the band with Weyl character, are given by the vector $| \psi^-\rangle$ with
components $a = (V^\prime-\epsilon S ) /(2\lambda_-)$, 
  $ b=-\epsilon / 2$, $c=1/2$, $d = -\epsilon a$, where $\epsilon\equiv sgn(X V^{\prime})$.

Using these eigenstates we can calculate the Berry connection vector,
$  {\bf A} \equiv - i \langle \psi^- | \bm{\nabla}_{\bf k} | \psi^- \rangle$,
\begin{eqnarray}
{\bf A} =
  {1 \over 2}  \bm{\nabla}_{\bf k} \phi_{\bf k}, \hskip 0.2 in 
  \phi_{\bf k} = \tan^{-1}\Bigl [{{Y} \over {X-\epsilon V^\prime}}\Bigr ].
  \label{phase-eq}
\end{eqnarray}
This implies for a line-integral on a circular contour centered around
any specific point on the nodal line, 
\begin{eqnarray}
  \oint {\bf A} \cdot d{\bf k} 
        =
        {1 \over 2} ( \phi_{\bf  k}(\alpha=2\pi)-\phi_{\bf  k}(\alpha=0)). 
        \label{bp-int}
\end{eqnarray}
The contour is in the plane
perpendicular to the $k_x$ direction and  $\alpha$ is the 
azimuthal angle which defines a point of the contour. See Appendix ~\ref{sec:AppendixB} for a more detailed description. 
The function $\phi_{\bf  k}(\alpha)$
has a singularity when the denominator of Eq.~(\ref{phase-eq}) vanishes.
This happens on a 2D surface 
and intersects the $k_y$-$k_z$ plane at the nodal line.
Any such closed contour around the nodal line crosses this surface twice,
with the value of $\phi_{\bf k}$ at 
each singularity contributing a value of $\pi$ to the integral, so we obtain,
$  \oint {\bf A} \cdot d{\bf k} = \pi$.

\section{Topological superconductivity and Majorana states}

\label{sec:interacting_model}
Having demonstrated that our model supports a Weyl nodal ring, we proceed by adding interactions
in order to study the exotic superconducting states that emerge in the ground state of the interacting model.

We consider a mean-field Bogoliubov-de-Gennes Hamiltonian of the form,

\begin{eqnarray}
\hat{H}^{\textmd{\tiny{BdG}}} &=&\frac{1}{2}\sum_{\mathbf{k}}
\begin{pmatrix}
c^\dagger_\mathbf{k} & c_\mathbf{-k} 
\end{pmatrix}
\hat {\mathcal {M}}
\begin{pmatrix}
c_\mathbf{k} & c^\dagger_\mathbf{-k} 
\end{pmatrix}, \nonumber \\
\hat {\mathcal {M}} & \equiv &
\begin{pmatrix}
\mathcal{H}_0(\mathbf{k})-\mu && \Delta \\
\Delta^\dagger && -\mathcal{H}^\intercal_0(\mathbf{-k})+\mu
\end{pmatrix}.
\label{eq:H_BdG}
\end{eqnarray} The matrix $\mathcal{H}_0(\mathbf{k})$ corresponds to Eq.~(\ref{eq:H0_nrs}), and the gap function $\Delta=\mathbb{I}_{2\times2} \otimes i\Delta_0\sigma_y$, where $\mathbb{I}_{2\times2}$ is the 
$2\times2$ identity matrix and $\sigma_y$ is a Pauli matrix.

\begin{figure}
    \begin{center}
      \includegraphics[width=1.\columnwidth]{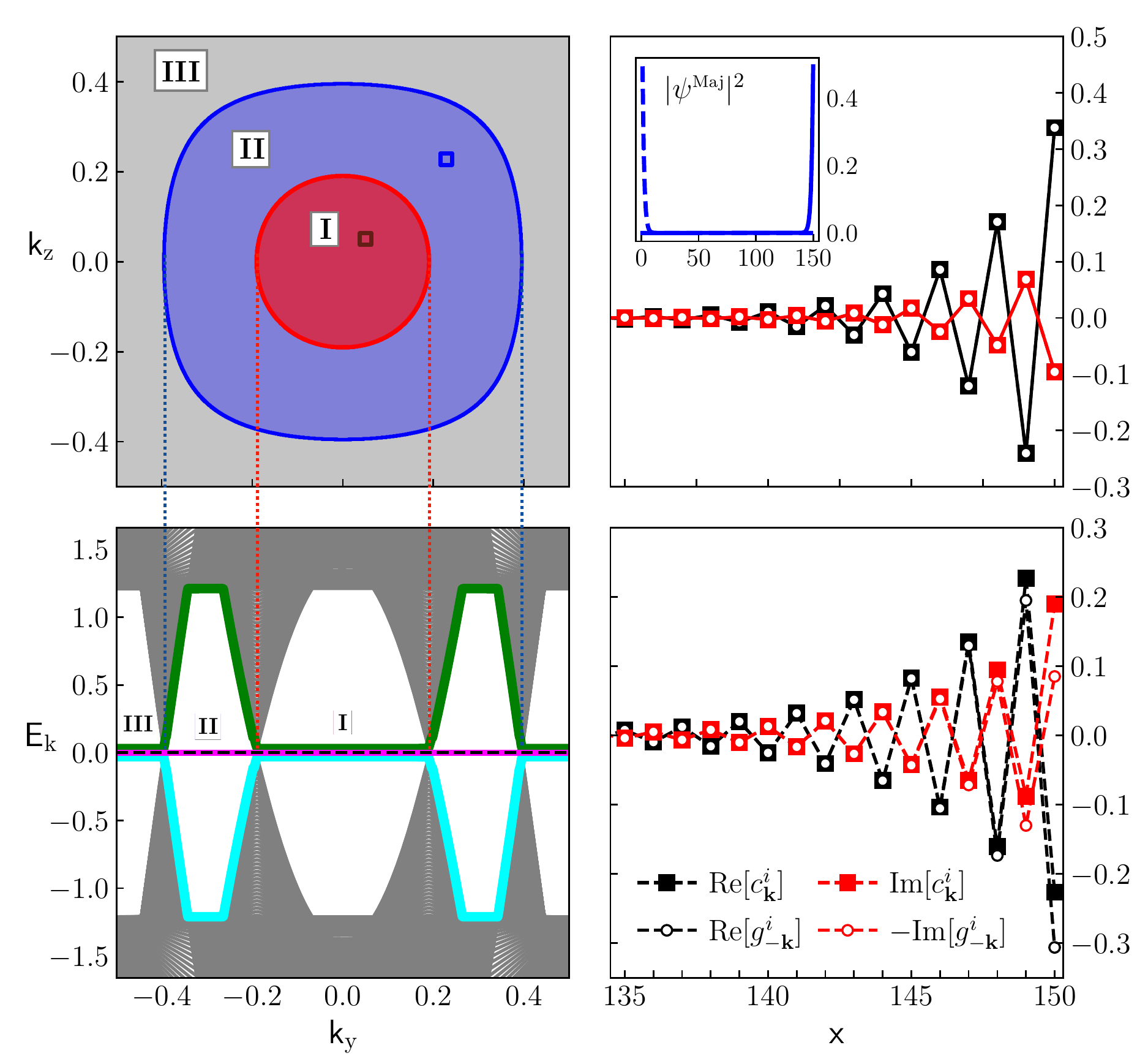}
    \end{center}
\caption{Nodal rings, surface spectrum and wavefunction amplitudes. (Upper left) Projected nodal ring in the $k_y$-$k_z$ plane, with
$\mu=0.2$, $\Delta=2.0$ ($k_y$ and $k_z$ are in units of $2\pi$, and the remaining parameters identical to Fig.~\ref{fig2}). Region I contains 4 drumhead states, region II, 2 annular Majorana surface states, and region III,
4 annular Majorana surface states. (Lower left) Surface spectrum along the $k_y$ axis. The vertical dashed lines are a guide to the eye for the region boundaries. We plot the surface states in color. 
(Upper right) Components of the wavefunction for a Majorana state 
at the momentum indicated by the blue box in the upper left panel. The inset shows the square modulus of the wavefunction for the two Majorana states at this momentum. 
(Lower right) Components of the wavefunction for a surface state at the momentum indicated by the red box in the upper left panel.\label{fig4}}
\end{figure}

We now solve the BdG equations in a slab geometry (with finite $x$ dimension). The presence of interactions causes the nodal
ring to split into two rings, which bound regions hosting different numbers of surface states with distinct topological characters (see upper left panel of
Fig.~\ref{fig4}). These surface states are dispersionless even in the presence of finite chemical potential (see lower left panel of Fig.~\ref{fig4}).

In order to characterize these surface states as Majorana states we examine the components of their wave functions. The Majorana states are defined by the property of 
their creation and annihilation operators, $\gamma_\mathbf{k}^\dagger=\gamma_\mathbf{-k}$. The Hamiltonian in Eq.~(\ref{eq:H_BdG}),
for the case of finite $x$ dimension, is diagonalized by Bogoliubov operators,
\begin{align}
& \gamma^{\dagger}_{k_\parallel} =  \sum_i a^i_{k_\parallel} c^{\dagger(\textmd{A})}_{i,k_\parallel}+b^i_{k_\parallel} c^{\dagger(\textmd{B})}_{i,k_\parallel} +c^i_{k_\parallel} c^{\dagger(\textmd{C})}_{i,k_\parallel} + d^i_{k_\parallel} c^{\dagger(\textmd{D})}_{i,k_\parallel} \notag\\
&+ e^i_{k_\parallel} c^{(\textmd{A})}_{i,-k_\parallel}+f^i_{k_\parallel} c^{(\textmd{B})}_{i,-k_\parallel} +g^i_{k_\parallel} c^{(\textmd{C})}_{i,-k_\parallel} + h^i_{k_\parallel} c^{(\textmd{D})}_{i,-k_\parallel},
\end{align}
where  $i$ labels the layer in the finite direction, and $k_\parallel=(k_y,k_z)$.
The Majorana condition $\gamma_\mathbf{k}^\dagger=\gamma_\mathbf{-k}$
implies  (up to an arbitrary phase),
\begin{eqnarray}
a^i_{k_\parallel}&=&(e^i_{-k_\parallel})^\ast , 
b^i_{k_\parallel}=(f^i_{-k_\parallel})^\ast \nonumber ,  \\
c^i_{k_\parallel}&=&(g^i_{-k_\parallel})^\ast ,
d^i_{k_\parallel}=(h^i_{-k_\parallel})^\ast. \label{eq:Maj_cond}
\end{eqnarray}

In the right column of Fig.~\ref{fig4} we plot the amplitudes $c^i_{k_\parallel}$, and $g^i_{-k_\parallel}$ of the wavefunction for a Majorana surface state (top) at $k_y/2\pi=k_z/2\pi=0.05$
and a drumhead state (bottom) at $k_y/2\pi=k_z/2\pi=0.228$. Both states are localized on the edge of the system, however only the Majorana state satisfies the condition of Eq.~(\ref{eq:Maj_cond}).
We plot only the amplitudes $c^i_{k_\parallel}$ and $g^i_{-k_\parallel}$, but we note that all components of the Majorana state satisfy Eq.~(\ref{eq:Maj_cond}).

\section{Summary and Conclusions}

Since its introduction, the SSH model has been one of the most important and well-studied models of condensed matter physics.
The model has served as a foundational description of many of the novel concepts that have since become central to the field of condensed matter.
In this paper we have introduced an analytically soluble 3D extension of
the SSH model which, for appropriate choices of its hopping parameters, describes well the band-structure of crystalline polyacetylene as obtained by DFT.
We observe that when a specific diagonal hopping is made sufficiently large a Weyl nodal ring forms in the band structure, whose projection bounds 
drumhead topological surface states. This intriguing state of 3D stacked SSH chains, in principle, can be realized
by applying high pressure on the crystallized polyacetylene or by doping or intercalation of atoms which can increase the effective 3D hopping amplitudes.
We find that both the $P2_1{/}a$ and the $P2_1{/}n$ structures, either of which may be stabilized under pressure, support a nodal ring state, and that this state is stable across a large parameter regime, 
which offers a broad window for experimental realizations. 
Starting from this Weyl nodal ring structure, by adding an interaction coupling electrons on carbon atoms of different sublattices within the same unit cell, we find topological superconductivity 
supporting both Bogoliubov-Weyl quasiparticles and annular Majorana surface states.
In addition to the fascinating physics this system displays, it also provides a new perspective on a well-known model, placing it again at the heart of modern condensed matter physics and reorienting it towards the new direction of exotic topological phenomena, including topological superconductivity. 

This work was supported in part by the Canada First Research Excellence Fund, and 
the U.S. National High Magnetic Field Laboratory, which is funded by NSF/
DMR-1644779 and the State of Florida.

\FloatBarrier
\appendix
\section{TB model for $P2_1{/}n$ structure}
\label{sec:AppendixA}
\begin{figure}[ht]
    \begin{center}
      \includegraphics[width=0.8\columnwidth]{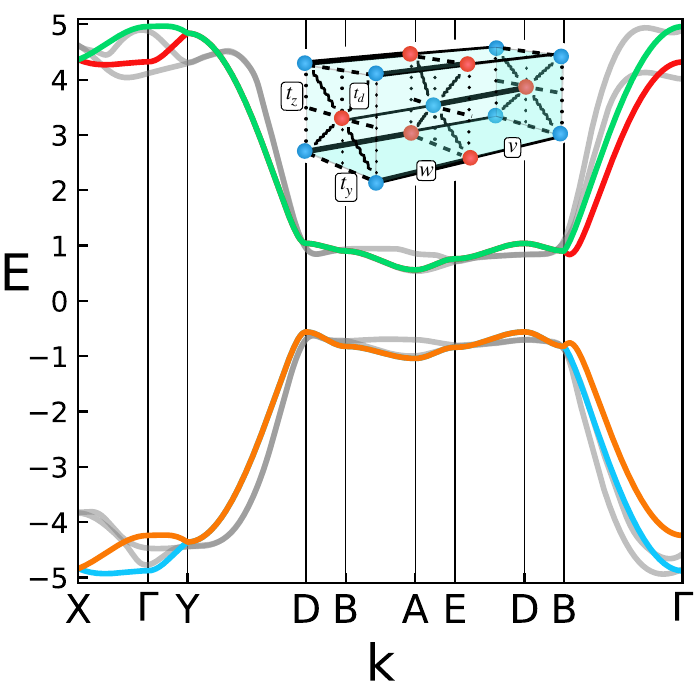}        
    \end{center}
    \caption{Unit cell and band structure for the model of $P2_1{/}n$ crystalline polyacetylene.          \label{struct_n}}
    \end{figure}

In Fig.~\ref{struct_n} we provide the unit cell for our TB model of the $P2_1{/}n$ structure of crystalline polyacetylene.
This model yields four bands of the form,
\begin{align}
E^\pm_1(\mathbf{k}) & = -2t_y\cos(k_ya_y)-2t_z\cos(k_za_z) \notag \\ 
					&\pm \sqrt{X^2+Y^2+V^2+2 \vert V \vert \vert X \vert} \notag \\
E^\pm_2(\mathbf{k}) & = -2t_y\cos(k_ya_y)-2t_z\cos(k_za_z) \notag \\
					&\pm \sqrt{X^2+Y^2+V^2-2 \vert V \vert \vert X \vert},
					\label{eq:bands}
\end{align}
which are plotted along with the DFT results (gray curves).

\section{Nodal line and Berry phase calculation}
\label{sec:AppendixB}
\begin{figure}
    \begin{center}
      \includegraphics[width=0.8\columnwidth]{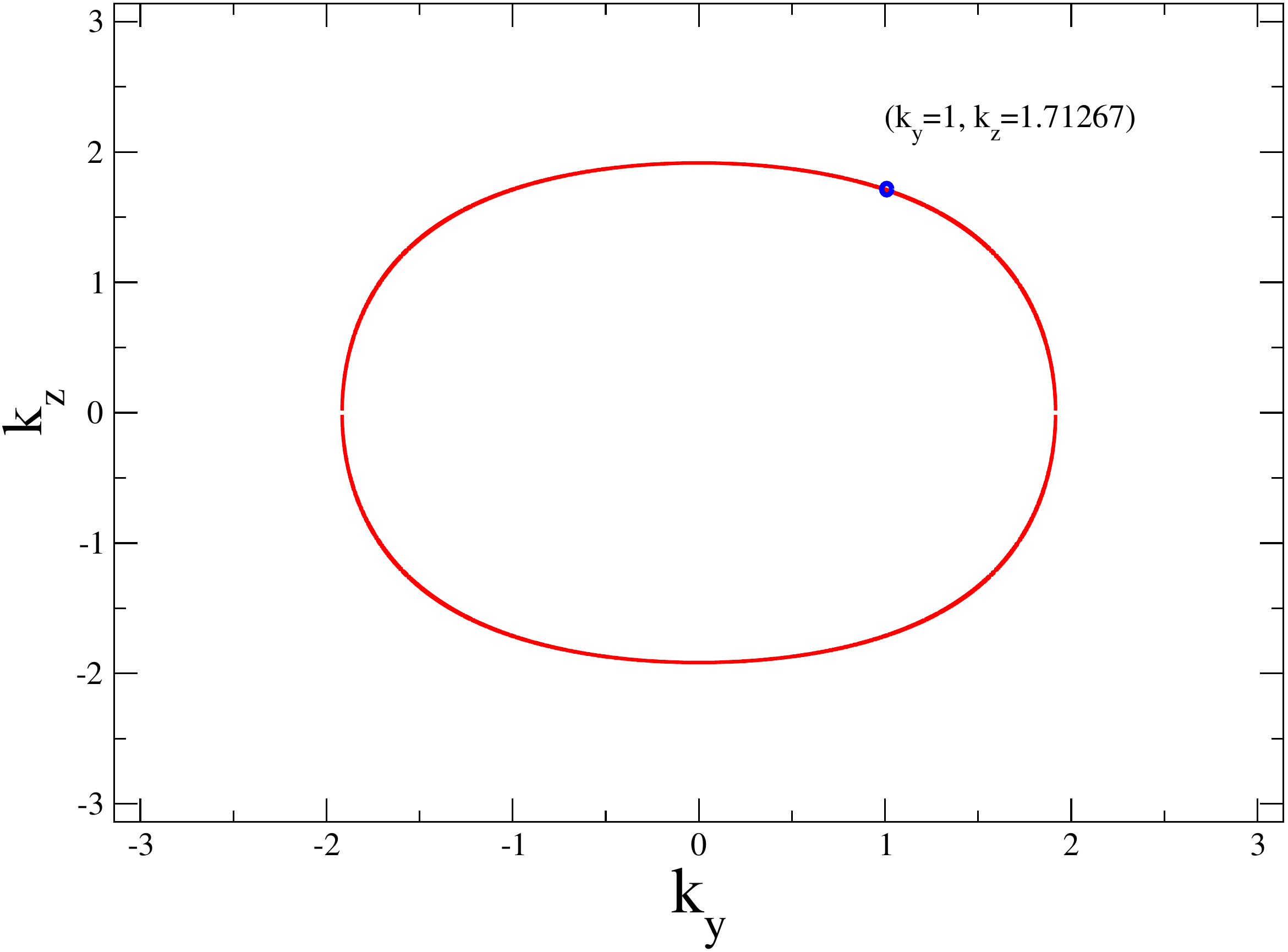}
    \end{center}
    \caption{An example of the nodal ring.          \label{nodal}}
    \end{figure}

Fig.~\ref{nodal} demonstrates an example of a nodal line on the $k_y$-$k_z$ plane, which is obtained
for a definite value of the ratio parameter $r$ defined by Eq.~(\ref{nodal_points}) of the main
manuscript.
The blue circle in the figure denotes a particular point on the nodal line, with coordinates
$(k_y,k_z)=(1,1.71267)$, that we have selected to demonstrate below how
the integral in Eq.~(\ref{bp-int}) of the main manuscript is evaluated.

\begin{figure}
  \begin{center}
            \subfigure[]{
              \includegraphics[width=0.835\columnwidth]{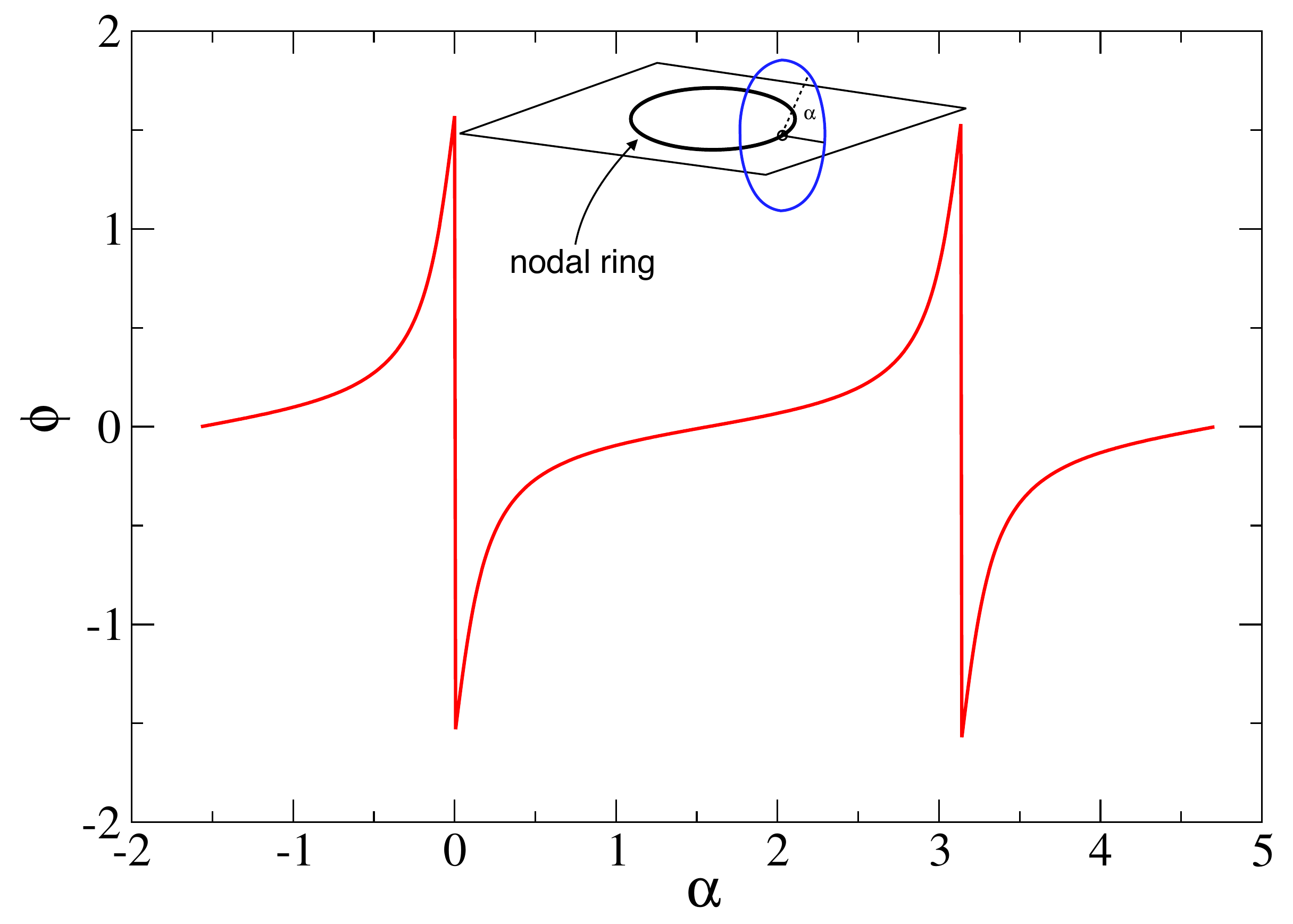}
            }
              \subfigure[]{
      \includegraphics[width=0.8\columnwidth]{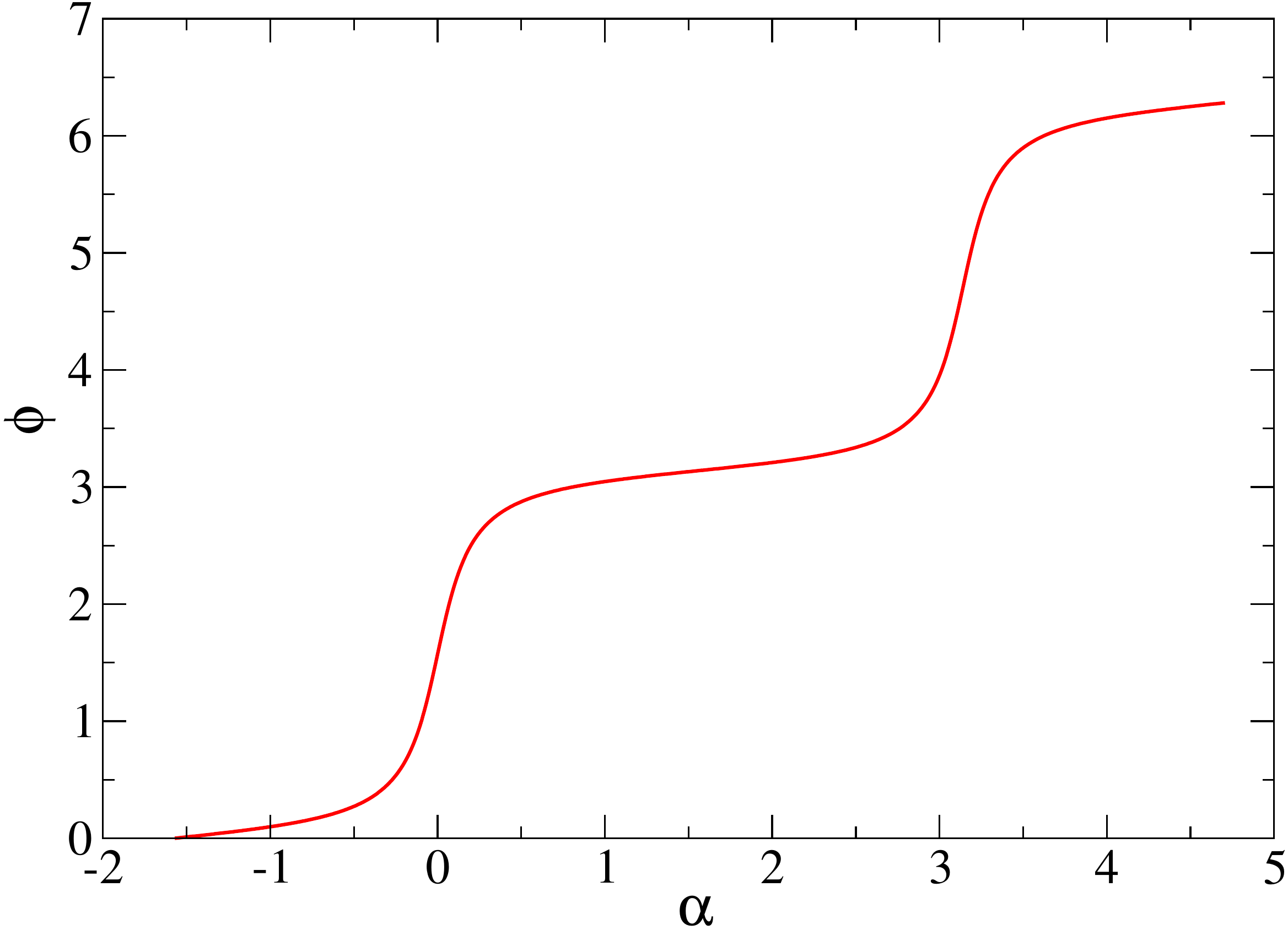}
}
      
    \end{center}
    \caption{(a) The function $\phi$ is plotted as a function
      of the angle $\alpha$, by taking its value in the interval
      $(-\pi/2,\pi/2]$. (b) The function $\phi$ is plotted as a function
    of the angle $\alpha$ by choosing the value, from the multitude
    of its values, that yields a continuous function.
      \label{phase}}

\end{figure}

After selecting a point on the nodal line,  which lies on the $k_y$-$k_z$ plane,
we define a circular  loop around the nodal line which lies
on a plane perpendicular to the nodal line. This loop is schematically
drawn in Fig.~\ref{phase} as an inset.
Then, we wish to calculate the line integral $
  \oint {\bf A} \cdot d{\bf k} $
around this loop. The instantaneous position on the loop is specified by the
angle $\alpha$ shown in the inset of Fig.~\ref{phase}.

In Fig.~\ref{phase} the function $\phi_{\bf  k}(\alpha)$ is plotted
as a function of $\alpha$ as extracted from Eq.~(\ref{phase-eq}) numerically.
However, we note that the function $\tan^{-1}$ is a multivalued function and
the function plotted in Fig.~\ref{phase}(a) only gives the value of the
function in the interval $[-\pi/2,\pi/2]$. This definition only works
correctly when the angle is needed in an interval of $\alpha$ which contains 
no singularity or branch cut. However, the function $\tan\phi_{\bf k}$
has a singularity when the denominator of Eq.~(\ref{phase-eq}) vanishes.
This happens on a 2D surface which is perpendicular to the $k_x$ direction
and intersects the $k_y$-$k_z$ plane at the nodal line.
So, the path crosses this surface twice. Since we are
looking for a smooth, i.e., continuous, transition of
the phase $\phi$ as a function of $\alpha$ we need to pick the
value (from the multitude of its possible values) of $\tan^{-1}$
that yields a continuous line. This implies that we need to
add $\pi$ for each of the singularities shown in Fig.~\ref{phase}(a).
Namely, we need to choose the solution which is plotted
in Fig.~\ref{phase}(b). Therefore, we obtain
\begin{eqnarray}
  \oint {\bf A} \cdot d{\bf k} = {1 \over 2} (\phi_{\bf k}(\alpha=2\pi)-
  \phi_{\bf k}(\alpha=0))
  = \pi.
\end{eqnarray}

\FloatBarrier


\begin{thebibliography}{36}%
\makeatletter
\providecommand \@ifxundefined [1]{%
 \@ifx{#1\undefined}
}%
\providecommand \@ifnum [1]{%
 \ifnum #1\expandafter \@firstoftwo
 \else \expandafter \@secondoftwo
 \fi
}%
\providecommand \@ifx [1]{%
 \ifx #1\expandafter \@firstoftwo
 \else \expandafter \@secondoftwo
 \fi
}%
\providecommand \natexlab [1]{#1}%
\providecommand \enquote  [1]{``#1''}%
\providecommand \bibnamefont  [1]{#1}%
\providecommand \bibfnamefont [1]{#1}%
\providecommand \citenamefont [1]{#1}%
\providecommand \href@noop [0]{\@secondoftwo}%
\providecommand \href [0]{\begingroup \@sanitize@url \@href}%
\providecommand \@href[1]{\@@startlink{#1}\@@href}%
\providecommand \@@href[1]{\endgroup#1\@@endlink}%
\providecommand \@sanitize@url [0]{\catcode `\\12\catcode `\$12\catcode
  `\&12\catcode `\#12\catcode `\^12\catcode `\_12\catcode `\%12\relax}%
\providecommand \@@startlink[1]{}%
\providecommand \@@endlink[0]{}%
\providecommand \url  [0]{\begingroup\@sanitize@url \@url }%
\providecommand \@url [1]{\endgroup\@href {#1}{\urlprefix }}%
\providecommand \urlprefix  [0]{URL }%
\providecommand \Eprint [0]{\href }%
\providecommand \doibase [0]{http://dx.doi.org/}%
\providecommand \selectlanguage [0]{\@gobble}%
\providecommand \bibinfo  [0]{\@secondoftwo}%
\providecommand \bibfield  [0]{\@secondoftwo}%
\providecommand \translation [1]{[#1]}%
\providecommand \BibitemOpen [0]{}%
\providecommand \bibitemStop [0]{}%
\providecommand \bibitemNoStop [0]{.\EOS\space}%
\providecommand \EOS [0]{\spacefactor3000\relax}%
\providecommand \BibitemShut  [1]{\csname bibitem#1\endcsname}%
\let\auto@bib@innerbib\@empty
%</preamble>
\bibitem [{\citenamefont {Su}\ \emph {et~al.}(1979)\citenamefont {Su},
  \citenamefont {Schrieffer},\ and\ \citenamefont {Heeger}}]{SSH1979}%
  \BibitemOpen
  \bibfield  {author} {\bibinfo {author} {\bibfnamefont {W.~P.}\ \bibnamefont
  {Su}}, \bibinfo {author} {\bibfnamefont {J.~R.}\ \bibnamefont {Schrieffer}},
  \ and\ \bibinfo {author} {\bibfnamefont {A.~J.}\ \bibnamefont {Heeger}},\
  }\href {\doibase 10.1103/PhysRevLett.42.1698} {\bibfield  {journal} {\bibinfo
   {journal} {Phys. Rev. Lett.}\ }\textbf {\bibinfo {volume} {42}},\ \bibinfo
  {pages} {1698} (\bibinfo {year} {1979})}\BibitemShut {NoStop}%
\bibitem [{\citenamefont {Heeger}\ \emph {et~al.}(1988)\citenamefont {Heeger},
  \citenamefont {Kivelson}, \citenamefont {Schrieffer},\ and\ \citenamefont
  {Su}}]{SSH_RMP}%
  \BibitemOpen
  \bibfield  {author} {\bibinfo {author} {\bibfnamefont {A.~J.}\ \bibnamefont
  {Heeger}}, \bibinfo {author} {\bibfnamefont {S.}~\bibnamefont {Kivelson}},
  \bibinfo {author} {\bibfnamefont {J.~R.}\ \bibnamefont {Schrieffer}}, \ and\
  \bibinfo {author} {\bibfnamefont {W.~P.}\ \bibnamefont {Su}},\ }\href
  {\doibase 10.1103/RevModPhys.60.781} {\bibfield  {journal} {\bibinfo
  {journal} {Rev. Mod. Phys.}\ }\textbf {\bibinfo {volume} {60}},\ \bibinfo
  {pages} {781} (\bibinfo {year} {1988})}\BibitemShut {NoStop}%
\bibitem [{\citenamefont {Rice}\ and\ \citenamefont {Mele}(1982)}]{Rice1982}%
  \BibitemOpen
  \bibfield  {author} {\bibinfo {author} {\bibfnamefont {M.~J.}\ \bibnamefont
  {Rice}}\ and\ \bibinfo {author} {\bibfnamefont {E.~J.}\ \bibnamefont
  {Mele}},\ }\href {\doibase 10.1103/PhysRevLett.49.1455} {\bibfield  {journal}
  {\bibinfo  {journal} {Phys. Rev. Lett.}\ }\textbf {\bibinfo {volume} {49}},\
  \bibinfo {pages} {1455} (\bibinfo {year} {1982})}\BibitemShut {NoStop}%
\bibitem [{\citenamefont {Zak}(1989)}]{Zak1989}%
  \BibitemOpen
  \bibfield  {author} {\bibinfo {author} {\bibfnamefont {J.}~\bibnamefont
  {Zak}},\ }\href {\doibase 10.1103/PhysRevLett.62.2747} {\bibfield  {journal}
  {\bibinfo  {journal} {Phys. Rev. Lett.}\ }\textbf {\bibinfo {volume} {62}},\
  \bibinfo {pages} {2747} (\bibinfo {year} {1989})}\BibitemShut {NoStop}%
\bibitem [{\citenamefont {Hasan}\ and\ \citenamefont {Kane}(2010)}]{Hasan2010}%
  \BibitemOpen
  \bibfield  {author} {\bibinfo {author} {\bibfnamefont {M.~Z.}\ \bibnamefont
  {Hasan}}\ and\ \bibinfo {author} {\bibfnamefont {C.~L.}\ \bibnamefont
  {Kane}},\ }\href {\doibase 10.1103/RevModPhys.82.3045} {\bibfield  {journal}
  {\bibinfo  {journal} {Rev. Mod. Phys.}\ }\textbf {\bibinfo {volume} {82}},\
  \bibinfo {pages} {3045} (\bibinfo {year} {2010})}\BibitemShut {NoStop}%
\bibitem [{\citenamefont {Armitage}\ \emph {et~al.}(2018)\citenamefont
  {Armitage}, \citenamefont {Mele},\ and\ \citenamefont
  {Vishwanath}}]{Armitage2018}%
  \BibitemOpen
  \bibfield  {author} {\bibinfo {author} {\bibfnamefont {N.~P.}\ \bibnamefont
  {Armitage}}, \bibinfo {author} {\bibfnamefont {E.~J.}\ \bibnamefont {Mele}},
  \ and\ \bibinfo {author} {\bibfnamefont {A.}~\bibnamefont {Vishwanath}},\
  }\href {\doibase 10.1103/RevModPhys.90.015001} {\bibfield  {journal}
  {\bibinfo  {journal} {Rev. Mod. Phys.}\ }\textbf {\bibinfo {volume} {90}},\
  \bibinfo {pages} {015001} (\bibinfo {year} {2018})}\BibitemShut {NoStop}%
\bibitem [{\citenamefont {Liu}\ \emph {et~al.}(2014)\citenamefont {Liu},
  \citenamefont {Zhou}, \citenamefont {Zhang}, \citenamefont {Wang},
  \citenamefont {Weng}, \citenamefont {Prabhakaran}, \citenamefont {Mo},
  \citenamefont {Shen}, \citenamefont {Fang}, \citenamefont {Dai},
  \citenamefont {Hussain},\ and\ \citenamefont {Chen}}]{Liu2014}%
  \BibitemOpen
  \bibfield  {author} {\bibinfo {author} {\bibfnamefont {Z.~K.}\ \bibnamefont
  {Liu}}, \bibinfo {author} {\bibfnamefont {B.}~\bibnamefont {Zhou}}, \bibinfo
  {author} {\bibfnamefont {Y.}~\bibnamefont {Zhang}}, \bibinfo {author}
  {\bibfnamefont {Z.~J.}\ \bibnamefont {Wang}}, \bibinfo {author}
  {\bibfnamefont {H.~M.}\ \bibnamefont {Weng}}, \bibinfo {author}
  {\bibfnamefont {D.}~\bibnamefont {Prabhakaran}}, \bibinfo {author}
  {\bibfnamefont {S.-K.}\ \bibnamefont {Mo}}, \bibinfo {author} {\bibfnamefont
  {Z.~X.}\ \bibnamefont {Shen}}, \bibinfo {author} {\bibfnamefont
  {Z.}~\bibnamefont {Fang}}, \bibinfo {author} {\bibfnamefont {X.}~\bibnamefont
  {Dai}}, \bibinfo {author} {\bibfnamefont {Z.}~\bibnamefont {Hussain}}, \ and\
  \bibinfo {author} {\bibfnamefont {Y.~L.}\ \bibnamefont {Chen}},\ }\href
  {\doibase 10.1126/science.1245085} {\bibfield  {journal} {\bibinfo  {journal}
  {Science}\ }\textbf {\bibinfo {volume} {343}},\ \bibinfo {pages} {864}
  (\bibinfo {year} {2014})}\BibitemShut {NoStop}%
\bibitem [{\citenamefont {Lu}\ \emph {et~al.}(2015)\citenamefont {Lu},
  \citenamefont {Wang}, \citenamefont {Ye}, \citenamefont {Ran}, \citenamefont
  {Fu}, \citenamefont {Joannopoulos},\ and\ \citenamefont {Solja{\v
  c}i{\'c}}}]{Lu2015}%
  \BibitemOpen
  \bibfield  {author} {\bibinfo {author} {\bibfnamefont {L.}~\bibnamefont
  {Lu}}, \bibinfo {author} {\bibfnamefont {Z.}~\bibnamefont {Wang}}, \bibinfo
  {author} {\bibfnamefont {D.}~\bibnamefont {Ye}}, \bibinfo {author}
  {\bibfnamefont {L.}~\bibnamefont {Ran}}, \bibinfo {author} {\bibfnamefont
  {L.}~\bibnamefont {Fu}}, \bibinfo {author} {\bibfnamefont {J.~D.}\
  \bibnamefont {Joannopoulos}}, \ and\ \bibinfo {author} {\bibfnamefont
  {M.}~\bibnamefont {Solja{\v c}i{\'c}}},\ }\href
  {https://science.sciencemag.org/content/349/6248/622} {\bibfield  {journal}
  {\bibinfo  {journal} {Science}\ }\textbf {\bibinfo {volume} {349}},\ \bibinfo
  {pages} {622} (\bibinfo {year} {2015})}\BibitemShut {NoStop}%
\bibitem [{\citenamefont {Lv}\ \emph {et~al.}(2015)\citenamefont {Lv},
  \citenamefont {Weng}, \citenamefont {Fu}, \citenamefont {Wang}, \citenamefont
  {Miao}, \citenamefont {Ma}, \citenamefont {Richard}, \citenamefont {Huang},
  \citenamefont {Zhao}, \citenamefont {Chen}, \citenamefont {Fang},
  \citenamefont {Dai}, \citenamefont {Qian},\ and\ \citenamefont
  {Ding}}]{Lv2015}%
  \BibitemOpen
  \bibfield  {author} {\bibinfo {author} {\bibfnamefont {B.~Q.}\ \bibnamefont
  {Lv}}, \bibinfo {author} {\bibfnamefont {H.~M.}\ \bibnamefont {Weng}},
  \bibinfo {author} {\bibfnamefont {B.~B.}\ \bibnamefont {Fu}}, \bibinfo
  {author} {\bibfnamefont {X.~P.}\ \bibnamefont {Wang}}, \bibinfo {author}
  {\bibfnamefont {H.}~\bibnamefont {Miao}}, \bibinfo {author} {\bibfnamefont
  {J.}~\bibnamefont {Ma}}, \bibinfo {author} {\bibfnamefont {P.}~\bibnamefont
  {Richard}}, \bibinfo {author} {\bibfnamefont {X.~C.}\ \bibnamefont {Huang}},
  \bibinfo {author} {\bibfnamefont {L.~X.}\ \bibnamefont {Zhao}}, \bibinfo
  {author} {\bibfnamefont {G.~F.}\ \bibnamefont {Chen}}, \bibinfo {author}
  {\bibfnamefont {Z.}~\bibnamefont {Fang}}, \bibinfo {author} {\bibfnamefont
  {X.}~\bibnamefont {Dai}}, \bibinfo {author} {\bibfnamefont {T.}~\bibnamefont
  {Qian}}, \ and\ \bibinfo {author} {\bibfnamefont {H.}~\bibnamefont {Ding}},\
  }\href {\doibase 10.1103/PhysRevX.5.031013} {\bibfield  {journal} {\bibinfo
  {journal} {Phys. Rev. X}\ }\textbf {\bibinfo {volume} {5}},\ \bibinfo {pages}
  {031013} (\bibinfo {year} {2015})}\BibitemShut {NoStop}%
\bibitem [{\citenamefont {Xu}\ \emph {et~al.}(2015)\citenamefont {Xu},
  \citenamefont {Belopolski}, \citenamefont {Alidoust}, \citenamefont
  {Neupane}, \citenamefont {Bian}, \citenamefont {Zhang}, \citenamefont
  {Sankar}, \citenamefont {Chang}, \citenamefont {Yuan}, \citenamefont {Lee},
  \citenamefont {Huang}, \citenamefont {Zheng}, \citenamefont {Ma},
  \citenamefont {Sanchez}, \citenamefont {Wang}, \citenamefont {Bansil},
  \citenamefont {Chou}, \citenamefont {Shibayev}, \citenamefont {Lin},
  \citenamefont {Jia},\ and\ \citenamefont {Hasan}}]{Xu2015}%
  \BibitemOpen
  \bibfield  {author} {\bibinfo {author} {\bibfnamefont {S.-Y.}\ \bibnamefont
  {Xu}}, \bibinfo {author} {\bibfnamefont {I.}~\bibnamefont {Belopolski}},
  \bibinfo {author} {\bibfnamefont {N.}~\bibnamefont {Alidoust}}, \bibinfo
  {author} {\bibfnamefont {M.}~\bibnamefont {Neupane}}, \bibinfo {author}
  {\bibfnamefont {G.}~\bibnamefont {Bian}}, \bibinfo {author} {\bibfnamefont
  {C.}~\bibnamefont {Zhang}}, \bibinfo {author} {\bibfnamefont
  {R.}~\bibnamefont {Sankar}}, \bibinfo {author} {\bibfnamefont
  {G.}~\bibnamefont {Chang}}, \bibinfo {author} {\bibfnamefont
  {Z.}~\bibnamefont {Yuan}}, \bibinfo {author} {\bibfnamefont {C.-C.}\
  \bibnamefont {Lee}}, \bibinfo {author} {\bibfnamefont {S.-M.}\ \bibnamefont
  {Huang}}, \bibinfo {author} {\bibfnamefont {H.}~\bibnamefont {Zheng}},
  \bibinfo {author} {\bibfnamefont {J.}~\bibnamefont {Ma}}, \bibinfo {author}
  {\bibfnamefont {D.~S.}\ \bibnamefont {Sanchez}}, \bibinfo {author}
  {\bibfnamefont {B.}~\bibnamefont {Wang}}, \bibinfo {author} {\bibfnamefont
  {A.}~\bibnamefont {Bansil}}, \bibinfo {author} {\bibfnamefont
  {F.}~\bibnamefont {Chou}}, \bibinfo {author} {\bibfnamefont {P.~P.}\
  \bibnamefont {Shibayev}}, \bibinfo {author} {\bibfnamefont {H.}~\bibnamefont
  {Lin}}, \bibinfo {author} {\bibfnamefont {S.}~\bibnamefont {Jia}}, \ and\
  \bibinfo {author} {\bibfnamefont {M.~Z.}\ \bibnamefont {Hasan}},\ }\href
  {\doibase 10.1126/science.aaa9297} {\bibfield  {journal} {\bibinfo  {journal}
  {Science}\ }\textbf {\bibinfo {volume} {349}},\ \bibinfo {pages} {613}
  (\bibinfo {year} {2015})}\BibitemShut {NoStop}%
\bibitem [{\citenamefont {Huang}\ \emph {et~al.}(2016)\citenamefont {Huang},
  \citenamefont {McCormick}, \citenamefont {Ochi}, \citenamefont {Zhao},
  \citenamefont {Suzuki}, \citenamefont {Arita}, \citenamefont {Wu},
  \citenamefont {Mou}, \citenamefont {Cao}, \citenamefont {Yan}, \citenamefont
  {Trivedi},\ and\ \citenamefont {Kaminski}}]{Huang2016}%
  \BibitemOpen
  \bibfield  {author} {\bibinfo {author} {\bibfnamefont {L.}~\bibnamefont
  {Huang}}, \bibinfo {author} {\bibfnamefont {T.~M.}\ \bibnamefont
  {McCormick}}, \bibinfo {author} {\bibfnamefont {M.}~\bibnamefont {Ochi}},
  \bibinfo {author} {\bibfnamefont {Z.}~\bibnamefont {Zhao}}, \bibinfo {author}
  {\bibfnamefont {M.-T.}\ \bibnamefont {Suzuki}}, \bibinfo {author}
  {\bibfnamefont {R.}~\bibnamefont {Arita}}, \bibinfo {author} {\bibfnamefont
  {Y.}~\bibnamefont {Wu}}, \bibinfo {author} {\bibfnamefont {D.}~\bibnamefont
  {Mou}}, \bibinfo {author} {\bibfnamefont {H.}~\bibnamefont {Cao}}, \bibinfo
  {author} {\bibfnamefont {J.}~\bibnamefont {Yan}}, \bibinfo {author}
  {\bibfnamefont {N.}~\bibnamefont {Trivedi}}, \ and\ \bibinfo {author}
  {\bibfnamefont {A.}~\bibnamefont {Kaminski}},\ }\href
  {https://doi.org/10.1038/nmat4685} {\bibfield  {journal} {\bibinfo  {journal}
  {Nature Materials}\ }\textbf {\bibinfo {volume} {15}},\ \bibinfo {pages}
  {1155} (\bibinfo {year} {2016})}\BibitemShut {NoStop}%
\bibitem [{\citenamefont {Jiang}\ \emph {et~al.}(2017)\citenamefont {Jiang},
  \citenamefont {Liu}, \citenamefont {Yang}, \citenamefont {Rajamathi},
  \citenamefont {Qi}, \citenamefont {Yang}, \citenamefont {Chen}, \citenamefont
  {Peng}, \citenamefont {Hwang}, \citenamefont {Sun}, \citenamefont {Mo},
  \citenamefont {Vobornik}, \citenamefont {Fujii}, \citenamefont {Parkin},
  \citenamefont {Felser}, \citenamefont {Yan},\ and\ \citenamefont
  {Chen}}]{Jiang2017}%
  \BibitemOpen
  \bibfield  {author} {\bibinfo {author} {\bibfnamefont {J.}~\bibnamefont
  {Jiang}}, \bibinfo {author} {\bibfnamefont {Y.}~\bibnamefont {Liu},
  \bibfnamefont {Z.~K.and~Sun}}, \bibinfo {author} {\bibfnamefont {H.~F.}\
  \bibnamefont {Yang}}, \bibinfo {author} {\bibfnamefont {C.~R.}\ \bibnamefont
  {Rajamathi}}, \bibinfo {author} {\bibfnamefont {Y.~P.}\ \bibnamefont {Qi}},
  \bibinfo {author} {\bibfnamefont {L.~X.}\ \bibnamefont {Yang}}, \bibinfo
  {author} {\bibfnamefont {C.}~\bibnamefont {Chen}}, \bibinfo {author}
  {\bibfnamefont {H.}~\bibnamefont {Peng}}, \bibinfo {author} {\bibfnamefont
  {C.-C.}\ \bibnamefont {Hwang}}, \bibinfo {author} {\bibfnamefont {S.~Z.}\
  \bibnamefont {Sun}}, \bibinfo {author} {\bibfnamefont {S.-K.}\ \bibnamefont
  {Mo}}, \bibinfo {author} {\bibfnamefont {I.}~\bibnamefont {Vobornik}},
  \bibinfo {author} {\bibfnamefont {J.}~\bibnamefont {Fujii}}, \bibinfo
  {author} {\bibfnamefont {S.~S.~P.}\ \bibnamefont {Parkin}}, \bibinfo {author}
  {\bibfnamefont {C.}~\bibnamefont {Felser}}, \bibinfo {author} {\bibfnamefont
  {B.~H.}\ \bibnamefont {Yan}}, \ and\ \bibinfo {author} {\bibfnamefont
  {Y.~L.}\ \bibnamefont {Chen}},\ }\href {\doibase 10.1038/ncomms13973}
  {\bibfield  {journal} {\bibinfo  {journal} {Nature Communications}\ }\textbf
  {\bibinfo {volume} {8}},\ \bibinfo {pages} {13973} (\bibinfo {year}
  {2017})}\BibitemShut {NoStop}%
\bibitem [{\citenamefont {Burkov}\ \emph {et~al.}(2011)\citenamefont {Burkov},
  \citenamefont {Hook},\ and\ \citenamefont
  {Balents}}]{burkov_topological_2011}%
  \BibitemOpen
  \bibfield  {author} {\bibinfo {author} {\bibfnamefont {A.~A.}\ \bibnamefont
  {Burkov}}, \bibinfo {author} {\bibfnamefont {M.~D.}\ \bibnamefont {Hook}}, \
  and\ \bibinfo {author} {\bibfnamefont {L.}~\bibnamefont {Balents}},\ }\href
  {\doibase 10.1103/PhysRevB.84.235126} {\bibfield  {journal} {\bibinfo
  {journal} {Physical Review B}\ }\textbf {\bibinfo {volume} {84}},\ \bibinfo
  {pages} {235126} (\bibinfo {year} {2011})}\BibitemShut {NoStop}%
\bibitem [{\citenamefont {Kim}\ \emph {et~al.}(2015)\citenamefont {Kim},
  \citenamefont {Wieder}, \citenamefont {Kane},\ and\ \citenamefont
  {Rappe}}]{kim_dirac_2015}%
  \BibitemOpen
  \bibfield  {author} {\bibinfo {author} {\bibfnamefont {Y.}~\bibnamefont
  {Kim}}, \bibinfo {author} {\bibfnamefont {B.~J.}\ \bibnamefont {Wieder}},
  \bibinfo {author} {\bibfnamefont {C.}~\bibnamefont {Kane}}, \ and\ \bibinfo
  {author} {\bibfnamefont {A.~M.}\ \bibnamefont {Rappe}},\ }\href
  {https://link.aps.org/doi/10.1103/PhysRevLett.115.036806} {\bibfield
  {journal} {\bibinfo  {journal} {Physical Review Letters}\ }\textbf {\bibinfo
  {volume} {115}},\ \bibinfo {pages} {036806} (\bibinfo {year}
  {2015})}\BibitemShut {NoStop}%
\bibitem [{\citenamefont {Chen}\ \emph {et~al.}(2015)\citenamefont {Chen},
  \citenamefont {Xie}, \citenamefont {Yang}, \citenamefont {Pan}, \citenamefont
  {Zhang}, \citenamefont {Cohen},\ and\ \citenamefont
  {Zhang}}]{chen_nanostructured_2015}%
  \BibitemOpen
  \bibfield  {author} {\bibinfo {author} {\bibfnamefont {Y.}~\bibnamefont
  {Chen}}, \bibinfo {author} {\bibfnamefont {Y.}~\bibnamefont {Xie}}, \bibinfo
  {author} {\bibfnamefont {S.~A.}\ \bibnamefont {Yang}}, \bibinfo {author}
  {\bibfnamefont {H.}~\bibnamefont {Pan}}, \bibinfo {author} {\bibfnamefont
  {F.}~\bibnamefont {Zhang}}, \bibinfo {author} {\bibfnamefont {M.~L.}\
  \bibnamefont {Cohen}}, \ and\ \bibinfo {author} {\bibfnamefont
  {S.}~\bibnamefont {Zhang}},\ }\href {\doibase 10.1021/acs.nanolett.5b02978}
  {\bibfield  {journal} {\bibinfo  {journal} {Nano Letters}\ }\textbf {\bibinfo
  {volume} {15}},\ \bibinfo {pages} {6974} (\bibinfo {year}
  {2015})}\BibitemShut {NoStop}%
\bibitem [{\citenamefont {Fang}\ \emph {et~al.}(2016)\citenamefont {Fang},
  \citenamefont {Weng}, \citenamefont {Dai},\ and\ \citenamefont
  {Fang}}]{fang_topological_2016}%
  \BibitemOpen
  \bibfield  {author} {\bibinfo {author} {\bibfnamefont {C.}~\bibnamefont
  {Fang}}, \bibinfo {author} {\bibfnamefont {H.}~\bibnamefont {Weng}}, \bibinfo
  {author} {\bibfnamefont {X.}~\bibnamefont {Dai}}, \ and\ \bibinfo {author}
  {\bibfnamefont {Z.}~\bibnamefont {Fang}},\ }\href
  {https://iopscience.iop.org/article/10.1088/1674-1056/25/11/117106}
  {\bibfield  {journal} {\bibinfo  {journal} {Chinese Physics B}\ }\textbf
  {\bibinfo {volume} {25}},\ \bibinfo {pages} {117106} (\bibinfo {year}
  {2016})}\BibitemShut {NoStop}%
\bibitem [{\citenamefont {Yu}\ \emph {et~al.}(2015)\citenamefont {Yu},
  \citenamefont {Weng}, \citenamefont {Fang}, \citenamefont {Dai},\ and\
  \citenamefont {Hu}}]{yu_topological_2015}%
  \BibitemOpen
  \bibfield  {author} {\bibinfo {author} {\bibfnamefont {R.}~\bibnamefont
  {Yu}}, \bibinfo {author} {\bibfnamefont {H.}~\bibnamefont {Weng}}, \bibinfo
  {author} {\bibfnamefont {Z.}~\bibnamefont {Fang}}, \bibinfo {author}
  {\bibfnamefont {X.}~\bibnamefont {Dai}}, \ and\ \bibinfo {author}
  {\bibfnamefont {X.}~\bibnamefont {Hu}},\ }\href {\doibase
  10.1103/PhysRevLett.115.036807} {\bibfield  {journal} {\bibinfo  {journal}
  {Physical Review Letters}\ }\textbf {\bibinfo {volume} {115}},\ \bibinfo
  {pages} {036807} (\bibinfo {year} {2015})}\BibitemShut {NoStop}%
\bibitem [{\citenamefont {Bian}\ \emph
  {et~al.}(2016{\natexlab{a}})\citenamefont {Bian}, \citenamefont {Chang},
  \citenamefont {Zheng}, \citenamefont {Velury}, \citenamefont {Xu},
  \citenamefont {Neupert}, \citenamefont {Chiu}, \citenamefont {Huang},
  \citenamefont {Sanchez}, \citenamefont {Belopolski}, \citenamefont
  {Alidoust}, \citenamefont {Chen}, \citenamefont {Chang}, \citenamefont
  {Bansil}, \citenamefont {Jeng}, \citenamefont {Lin},\ and\ \citenamefont
  {Hasan}}]{bian_drumhead_2016}%
  \BibitemOpen
  \bibfield  {author} {\bibinfo {author} {\bibfnamefont {G.}~\bibnamefont
  {Bian}}, \bibinfo {author} {\bibfnamefont {T.-R.}\ \bibnamefont {Chang}},
  \bibinfo {author} {\bibfnamefont {H.}~\bibnamefont {Zheng}}, \bibinfo
  {author} {\bibfnamefont {S.}~\bibnamefont {Velury}}, \bibinfo {author}
  {\bibfnamefont {S.-Y.}\ \bibnamefont {Xu}}, \bibinfo {author} {\bibfnamefont
  {T.}~\bibnamefont {Neupert}}, \bibinfo {author} {\bibfnamefont {C.-K.}\
  \bibnamefont {Chiu}}, \bibinfo {author} {\bibfnamefont {S.-M.}\ \bibnamefont
  {Huang}}, \bibinfo {author} {\bibfnamefont {D.~S.}\ \bibnamefont {Sanchez}},
  \bibinfo {author} {\bibfnamefont {I.}~\bibnamefont {Belopolski}}, \bibinfo
  {author} {\bibfnamefont {N.}~\bibnamefont {Alidoust}}, \bibinfo {author}
  {\bibfnamefont {P.-J.}\ \bibnamefont {Chen}}, \bibinfo {author}
  {\bibfnamefont {G.}~\bibnamefont {Chang}}, \bibinfo {author} {\bibfnamefont
  {A.}~\bibnamefont {Bansil}}, \bibinfo {author} {\bibfnamefont {H.-T.}\
  \bibnamefont {Jeng}}, \bibinfo {author} {\bibfnamefont {H.}~\bibnamefont
  {Lin}}, \ and\ \bibinfo {author} {\bibfnamefont {M.~Z.}\ \bibnamefont
  {Hasan}},\ }\href {https://link.aps.org/doi/10.1103/PhysRevB.93.121113}
  {\bibfield  {journal} {\bibinfo  {journal} {Physical Review B}\ }\textbf
  {\bibinfo {volume} {93}},\ \bibinfo {pages} {121113} (\bibinfo {year}
  {2016}{\natexlab{a}})}\BibitemShut {NoStop}%
\bibitem [{\citenamefont {Bian}\ \emph
  {et~al.}(2016{\natexlab{b}})\citenamefont {Bian}, \citenamefont {Chang},
  \citenamefont {Sankar}, \citenamefont {Xu}, \citenamefont {Zheng},
  \citenamefont {Neupert}, \citenamefont {Chiu}, \citenamefont {Huang},
  \citenamefont {Chang}, \citenamefont {Belopolski}, \citenamefont {Sanchez},
  \citenamefont {Neupane}, \citenamefont {Alidoust}, \citenamefont {Liu},
  \citenamefont {Wang}, \citenamefont {Lee}, \citenamefont {Jeng},
  \citenamefont {Zhang}, \citenamefont {Yuan}, \citenamefont {Jia},
  \citenamefont {Bansil}, \citenamefont {Chou}, \citenamefont {Lin},\ and\
  \citenamefont {Hasan}}]{Bian2016}%
  \BibitemOpen
  \bibfield  {author} {\bibinfo {author} {\bibfnamefont {G.}~\bibnamefont
  {Bian}}, \bibinfo {author} {\bibfnamefont {T.-R.}\ \bibnamefont {Chang}},
  \bibinfo {author} {\bibfnamefont {R.}~\bibnamefont {Sankar}}, \bibinfo
  {author} {\bibfnamefont {S.-Y.}\ \bibnamefont {Xu}}, \bibinfo {author}
  {\bibfnamefont {H.}~\bibnamefont {Zheng}}, \bibinfo {author} {\bibfnamefont
  {T.}~\bibnamefont {Neupert}}, \bibinfo {author} {\bibfnamefont {C.-K.}\
  \bibnamefont {Chiu}}, \bibinfo {author} {\bibfnamefont {S.-M.}\ \bibnamefont
  {Huang}}, \bibinfo {author} {\bibfnamefont {G.}~\bibnamefont {Chang}},
  \bibinfo {author} {\bibfnamefont {I.}~\bibnamefont {Belopolski}}, \bibinfo
  {author} {\bibfnamefont {D.~S.}\ \bibnamefont {Sanchez}}, \bibinfo {author}
  {\bibfnamefont {M.}~\bibnamefont {Neupane}}, \bibinfo {author} {\bibfnamefont
  {N.}~\bibnamefont {Alidoust}}, \bibinfo {author} {\bibfnamefont
  {C.}~\bibnamefont {Liu}}, \bibinfo {author} {\bibfnamefont {B.}~\bibnamefont
  {Wang}}, \bibinfo {author} {\bibfnamefont {C.-C.}\ \bibnamefont {Lee}},
  \bibinfo {author} {\bibfnamefont {H.-T.}\ \bibnamefont {Jeng}}, \bibinfo
  {author} {\bibfnamefont {C.}~\bibnamefont {Zhang}}, \bibinfo {author}
  {\bibfnamefont {Z.}~\bibnamefont {Yuan}}, \bibinfo {author} {\bibfnamefont
  {S.}~\bibnamefont {Jia}}, \bibinfo {author} {\bibfnamefont {A.}~\bibnamefont
  {Bansil}}, \bibinfo {author} {\bibfnamefont {F.}~\bibnamefont {Chou}},
  \bibinfo {author} {\bibfnamefont {H.}~\bibnamefont {Lin}}, \ and\ \bibinfo
  {author} {\bibfnamefont {M.~Z.}\ \bibnamefont {Hasan}},\ }\href
  {https://doi.org/10.1038/ncomms10556} {\bibfield  {journal} {\bibinfo
  {journal} {Nature Communications}\ }\textbf {\bibinfo {volume} {7}},\
  \bibinfo {pages} {10556} (\bibinfo {year} {2016}{\natexlab{b}})}\BibitemShut
  {NoStop}%
\bibitem [{\citenamefont {Atala}\ \emph {et~al.}(2013)\citenamefont {Atala},
  \citenamefont {Aidelsburger}, \citenamefont {Barreiro}, \citenamefont
  {Abanin}, \citenamefont {Kitagawa}, \citenamefont {Demler},\ and\
  \citenamefont {Bloch}}]{Atala2013}%
  \BibitemOpen
  \bibfield  {author} {\bibinfo {author} {\bibfnamefont {M.}~\bibnamefont
  {Atala}}, \bibinfo {author} {\bibfnamefont {M.}~\bibnamefont {Aidelsburger}},
  \bibinfo {author} {\bibfnamefont {J.~T.}\ \bibnamefont {Barreiro}}, \bibinfo
  {author} {\bibfnamefont {D.}~\bibnamefont {Abanin}}, \bibinfo {author}
  {\bibfnamefont {T.}~\bibnamefont {Kitagawa}}, \bibinfo {author}
  {\bibfnamefont {E.}~\bibnamefont {Demler}}, \ and\ \bibinfo {author}
  {\bibfnamefont {I.}~\bibnamefont {Bloch}},\ }\href
  {https://doi.org/10.1038/nphys2790} {\bibfield  {journal} {\bibinfo
  {journal} {Nature Physics}\ }\textbf {\bibinfo {volume} {9}},\ \bibinfo
  {pages} {795} (\bibinfo {year} {2013})}\BibitemShut {NoStop}%
\bibitem [{\citenamefont {Cooper}\ \emph {et~al.}(2019)\citenamefont {Cooper},
  \citenamefont {Dalibard},\ and\ \citenamefont {Spielman}}]{Cooper2019}%
  \BibitemOpen
  \bibfield  {author} {\bibinfo {author} {\bibfnamefont {N.~R.}\ \bibnamefont
  {Cooper}}, \bibinfo {author} {\bibfnamefont {J.}~\bibnamefont {Dalibard}}, \
  and\ \bibinfo {author} {\bibfnamefont {I.~B.}\ \bibnamefont {Spielman}},\
  }\href {\doibase 10.1103/RevModPhys.91.015005} {\bibfield  {journal}
  {\bibinfo  {journal} {Rev. Mod. Phys.}\ }\textbf {\bibinfo {volume} {91}},\
  \bibinfo {pages} {015005} (\bibinfo {year} {2019})}\BibitemShut {NoStop}%
\bibitem [{\citenamefont {Wang}\ \emph {et~al.}(2021)\citenamefont {Wang},
  \citenamefont {Cheng}, \citenamefont {Wang}, \citenamefont {Zhang},
  \citenamefont {Lu}, \citenamefont {Yi}, \citenamefont {Niu}, \citenamefont
  {Deng}, \citenamefont {Liu}, \citenamefont {Chen},\ and\ \citenamefont
  {Pan}}]{wang_realization_2021}%
  \BibitemOpen
  \bibfield  {author} {\bibinfo {author} {\bibfnamefont {Z.-Y.}\ \bibnamefont
  {Wang}}, \bibinfo {author} {\bibfnamefont {X.-C.}\ \bibnamefont {Cheng}},
  \bibinfo {author} {\bibfnamefont {B.-Z.}\ \bibnamefont {Wang}}, \bibinfo
  {author} {\bibfnamefont {J.-Y.}\ \bibnamefont {Zhang}}, \bibinfo {author}
  {\bibfnamefont {Y.-H.}\ \bibnamefont {Lu}}, \bibinfo {author} {\bibfnamefont
  {C.-R.}\ \bibnamefont {Yi}}, \bibinfo {author} {\bibfnamefont
  {S.}~\bibnamefont {Niu}}, \bibinfo {author} {\bibfnamefont {Y.}~\bibnamefont
  {Deng}}, \bibinfo {author} {\bibfnamefont {X.-J.}\ \bibnamefont {Liu}},
  \bibinfo {author} {\bibfnamefont {S.}~\bibnamefont {Chen}}, \ and\ \bibinfo
  {author} {\bibfnamefont {J.-W.}\ \bibnamefont {Pan}},\ }\href
  {https://science.sciencemag.org/content/372/6539/271} {\bibfield  {journal}
  {\bibinfo  {journal} {Science}\ }\textbf {\bibinfo {volume} {372}},\ \bibinfo
  {pages} {271} (\bibinfo {year} {2021})}\BibitemShut {NoStop}%
\bibitem [{\citenamefont {Song}\ \emph {et~al.}(2019)\citenamefont {Song},
  \citenamefont {He}, \citenamefont {Niu}, \citenamefont {Zhang}, \citenamefont
  {Ren}, \citenamefont {Liu},\ and\ \citenamefont
  {Jo}}]{song_observation_2019}%
  \BibitemOpen
  \bibfield  {author} {\bibinfo {author} {\bibfnamefont {B.}~\bibnamefont
  {Song}}, \bibinfo {author} {\bibfnamefont {C.}~\bibnamefont {He}}, \bibinfo
  {author} {\bibfnamefont {S.}~\bibnamefont {Niu}}, \bibinfo {author}
  {\bibfnamefont {L.}~\bibnamefont {Zhang}}, \bibinfo {author} {\bibfnamefont
  {Z.}~\bibnamefont {Ren}}, \bibinfo {author} {\bibfnamefont {X.-J.}\
  \bibnamefont {Liu}}, \ and\ \bibinfo {author} {\bibfnamefont {G.-B.}\
  \bibnamefont {Jo}},\ }\href {\doibase 10.1038/s41567-019-0564-y} {\bibfield
  {journal} {\bibinfo  {journal} {Nature Physics}\ }\textbf {\bibinfo {volume}
  {15}},\ \bibinfo {pages} {911} (\bibinfo {year} {2019})}\BibitemShut
  {NoStop}%
\bibitem [{\citenamefont {Meng}\ and\ \citenamefont
  {Balents}(2012)}]{Balents_2012}%
  \BibitemOpen
  \bibfield  {author} {\bibinfo {author} {\bibfnamefont {T.}~\bibnamefont
  {Meng}}\ and\ \bibinfo {author} {\bibfnamefont {L.}~\bibnamefont {Balents}},\
  }\href {\doibase 10.1103/PhysRevB.86.054504} {\bibfield  {journal} {\bibinfo
  {journal} {Physical Review B}\ }\textbf {\bibinfo {volume} {86}},\ \bibinfo
  {pages} {054504} (\bibinfo {year} {2012})}\BibitemShut {NoStop}%
\bibitem [{\citenamefont {Cho}\ \emph {et~al.}(2012)\citenamefont {Cho},
  \citenamefont {Bardarson}, \citenamefont {Lu},\ and\ \citenamefont
  {Moore}}]{Cho2012}%
  \BibitemOpen
  \bibfield  {author} {\bibinfo {author} {\bibfnamefont {G.~Y.}\ \bibnamefont
  {Cho}}, \bibinfo {author} {\bibfnamefont {J.~H.}\ \bibnamefont {Bardarson}},
  \bibinfo {author} {\bibfnamefont {Y.-M.}\ \bibnamefont {Lu}}, \ and\ \bibinfo
  {author} {\bibfnamefont {J.~E.}\ \bibnamefont {Moore}},\ }\href
  {https://link.aps.org/doi/10.1103/PhysRevB.86.214514} {\bibfield  {journal}
  {\bibinfo  {journal} {Phys. Rev. B}\ }\textbf {\bibinfo {volume} {86}},\
  \bibinfo {pages} {214514} (\bibinfo {year} {2012})}\BibitemShut {NoStop}%
\bibitem [{\citenamefont {Bednik}\ \emph {et~al.}(2015)\citenamefont {Bednik},
  \citenamefont {Zyuzin},\ and\ \citenamefont {Burkov}}]{Burkov_2015}%
  \BibitemOpen
  \bibfield  {author} {\bibinfo {author} {\bibfnamefont {G.}~\bibnamefont
  {Bednik}}, \bibinfo {author} {\bibfnamefont {A.~A.}\ \bibnamefont {Zyuzin}},
  \ and\ \bibinfo {author} {\bibfnamefont {A.~A.}\ \bibnamefont {Burkov}},\
  }\href {\doibase 10.1103/PhysRevB.92.035153} {\bibfield  {journal} {\bibinfo
  {journal} {Physical Review B}\ }\textbf {\bibinfo {volume} {92}},\ \bibinfo
  {pages} {035153} (\bibinfo {year} {2015})}\BibitemShut {NoStop}%
\bibitem [{\citenamefont {Li}\ and\ \citenamefont
  {Haldane}(2018)}]{Haldane_2018}%
  \BibitemOpen
  \bibfield  {author} {\bibinfo {author} {\bibfnamefont {Y.}~\bibnamefont
  {Li}}\ and\ \bibinfo {author} {\bibfnamefont {F.}~\bibnamefont {Haldane}},\
  }\href {\doibase 10.1103/PhysRevLett.120.067003} {\bibfield  {journal}
  {\bibinfo  {journal} {Physical Review Letters}\ }\textbf {\bibinfo {volume}
  {120}},\ \bibinfo {pages} {067003} (\bibinfo {year} {2018})}\BibitemShut
  {NoStop}%
\bibitem [{\citenamefont {Kobayashi}\ and\ \citenamefont
  {Sato}(2015)}]{kobayashi_topological_2015}%
  \BibitemOpen
  \bibfield  {author} {\bibinfo {author} {\bibfnamefont {S.}~\bibnamefont
  {Kobayashi}}\ and\ \bibinfo {author} {\bibfnamefont {M.}~\bibnamefont
  {Sato}},\ }\href {\doibase 10.1103/PhysRevLett.115.187001} {\bibfield
  {journal} {\bibinfo  {journal} {Physical Review Letters}\ }\textbf {\bibinfo
  {volume} {115}},\ \bibinfo {pages} {187001} (\bibinfo {year}
  {2015})}\BibitemShut {NoStop}%
\bibitem [{\citenamefont {Alidoust}\ \emph {et~al.}(2017)\citenamefont
  {Alidoust}, \citenamefont {Halterman},\ and\ \citenamefont
  {Zyuzin}}]{Alidoust2017}%
  \BibitemOpen
  \bibfield  {author} {\bibinfo {author} {\bibfnamefont {M.}~\bibnamefont
  {Alidoust}}, \bibinfo {author} {\bibfnamefont {K.}~\bibnamefont {Halterman}},
  \ and\ \bibinfo {author} {\bibfnamefont {A.~A.}\ \bibnamefont {Zyuzin}},\
  }\href {\doibase 10.1103/PhysRevB.95.155124} {\bibfield  {journal} {\bibinfo
  {journal} {Phys. Rev. B}\ }\textbf {\bibinfo {volume} {95}},\ \bibinfo
  {pages} {155124} (\bibinfo {year} {2017})}\BibitemShut {NoStop}%
\bibitem [{\citenamefont {Rosenberg}\ \emph {et~al.}(2019)\citenamefont
  {Rosenberg}, \citenamefont {Aryal},\ and\ \citenamefont
  {Manousakis}}]{2DWeyl_AFQMC}%
  \BibitemOpen
  \bibfield  {author} {\bibinfo {author} {\bibfnamefont {P.}~\bibnamefont
  {Rosenberg}}, \bibinfo {author} {\bibfnamefont {N.}~\bibnamefont {Aryal}}, \
  and\ \bibinfo {author} {\bibfnamefont {E.}~\bibnamefont {Manousakis}},\
  }\href {\doibase 10.1103/PhysRevB.100.104522} {\bibfield  {journal} {\bibinfo
   {journal} {Phys. Rev. B}\ }\textbf {\bibinfo {volume} {100}},\ \bibinfo
  {pages} {104522} (\bibinfo {year} {2019})}\BibitemShut {NoStop}%
\bibitem [{\citenamefont {Nandkishore}(2016)}]{nandkishore_weyl_2016}%
  \BibitemOpen
  \bibfield  {author} {\bibinfo {author} {\bibfnamefont {R.}~\bibnamefont
  {Nandkishore}},\ }\href {\doibase 10.1103/PhysRevB.93.020506} {\bibfield
  {journal} {\bibinfo  {journal} {Physical Review B}\ }\textbf {\bibinfo
  {volume} {93}},\ \bibinfo {pages} {020506} (\bibinfo {year}
  {2016})}\BibitemShut {NoStop}%
\bibitem [{\citenamefont {Wang}\ and\ \citenamefont
  {Nandkishore}(2017)}]{wang_topological_2017}%
  \BibitemOpen
  \bibfield  {author} {\bibinfo {author} {\bibfnamefont {Y.}~\bibnamefont
  {Wang}}\ and\ \bibinfo {author} {\bibfnamefont {R.~M.}\ \bibnamefont
  {Nandkishore}},\ }\href {\doibase 10.1103/PhysRevB.95.060506} {\bibfield
  {journal} {\bibinfo  {journal} {Physical Review B}\ }\textbf {\bibinfo
  {volume} {95}},\ \bibinfo {pages} {060506} (\bibinfo {year}
  {2017})}\BibitemShut {NoStop}%
\bibitem [{\citenamefont {Fu}\ \emph {et~al.}(2020)\citenamefont {Fu},
  \citenamefont {Liu},\ and\ \citenamefont {Wu}}]{fu_transport_2020}%
  \BibitemOpen
  \bibfield  {author} {\bibinfo {author} {\bibfnamefont {P.-H.}\ \bibnamefont
  {Fu}}, \bibinfo {author} {\bibfnamefont {J.-F.}\ \bibnamefont {Liu}}, \ and\
  \bibinfo {author} {\bibfnamefont {J.}~\bibnamefont {Wu}},\ }\href
  {https://link.aps.org/doi/10.1103/PhysRevB.102.075430} {\bibfield  {journal}
  {\bibinfo  {journal} {Physical Review B}\ }\textbf {\bibinfo {volume}
  {102}},\ \bibinfo {pages} {075430} (\bibinfo {year} {2020})}\BibitemShut
  {NoStop}%
\bibitem [{\citenamefont {Faraei}\ and\ \citenamefont
  {Jafari}(2019)}]{faraei_induced_2019}%
  \BibitemOpen
  \bibfield  {author} {\bibinfo {author} {\bibfnamefont {Z.}~\bibnamefont
  {Faraei}}\ and\ \bibinfo {author} {\bibfnamefont {S.~A.}\ \bibnamefont
  {Jafari}},\ }\href {\doibase 10.1103/PhysRevB.100.035447} {\bibfield
  {journal} {\bibinfo  {journal} {Phys. Rev. B}\ }\textbf {\bibinfo {volume}
  {100}},\ \bibinfo {pages} {035447} (\bibinfo {year} {2019})}\BibitemShut
  {NoStop}%
\bibitem [{\citenamefont {Vogl}\ and\ \citenamefont
  {Campbell}(1990)}]{Vogl1990}%
  \BibitemOpen
  \bibfield  {author} {\bibinfo {author} {\bibfnamefont {P.}~\bibnamefont
  {Vogl}}\ and\ \bibinfo {author} {\bibfnamefont {D.~K.}\ \bibnamefont
  {Campbell}},\ }\href {\doibase 10.1103/PhysRevB.41.12797} {\bibfield
  {journal} {\bibinfo  {journal} {Phys. Rev. B}\ }\textbf {\bibinfo {volume}
  {41}},\ \bibinfo {pages} {12797} (\bibinfo {year} {1990})}\BibitemShut
  {NoStop}%
\bibitem [{\citenamefont {Chan}\ \emph {et~al.}(2016)\citenamefont {Chan},
  \citenamefont {Chiu}, \citenamefont {Chou},\ and\ \citenamefont
  {Schnyder}}]{chan_ca_2016}%
  \BibitemOpen
  \bibfield  {author} {\bibinfo {author} {\bibfnamefont {Y.-H.}\ \bibnamefont
  {Chan}}, \bibinfo {author} {\bibfnamefont {C.-K.}\ \bibnamefont {Chiu}},
  \bibinfo {author} {\bibfnamefont {M.~Y.}\ \bibnamefont {Chou}}, \ and\
  \bibinfo {author} {\bibfnamefont {A.~P.}\ \bibnamefont {Schnyder}},\ }\href
  {\doibase 10.1103/PhysRevB.93.205132} {\bibfield  {journal} {\bibinfo
  {journal} {Physical Review B}\ }\textbf {\bibinfo {volume} {93}},\ \bibinfo
  {pages} {205132} (\bibinfo {year} {2016})}\BibitemShut {NoStop}%
\end{thebibliography}
\end{document}